\documentclass[lettersize,journal]{IEEEtran}
\usepackage{amsmath,amsfonts}
\usepackage{algorithm}
\usepackage{algpseudocode}
\usepackage{array}
\usepackage[caption=false,font=normalsize,labelfont=sf,textfont=sf]{subfig}
\usepackage{textcomp}
\usepackage{stfloats}
\usepackage{url}
\usepackage{verbatim}
\usepackage{graphicx}
\usepackage{float} 
\usepackage{booktabs} 
\usepackage{scrextend}
\usepackage{amssymb}
\usepackage{multirow}
\usepackage{amsthm}   
\usepackage{pifont}
\usepackage[table]{xcolor}
\definecolor{lightgreen}{RGB}{105, 195, 105} 
\usepackage[style=ieee,backend=bibtex]{biblatex}
\usepackage[
pdfauthor={derajan},
pdftitle={How to do this},
pdfstartview=XYZ,
bookmarks=true,
colorlinks=true,
linkcolor=blue,
urlcolor=blue,
citecolor=blue,
pdftex,
linktocpage=true,   
hyperindex=true
]{hyperref}
\makeatother
\begin{document}

\title{Joint Lossless Compression and Steganography for Medical Images via Large Language Models}

\author{Pengcheng Zheng, Xiaorong Pu, Kecheng Chen, \IEEEmembership{Graduate Student Member, IEEE}, Jiaxin Huang, \IEEEmembership{Graduate Student Member, IEEE}, Meng Yang, Bai Feng, Yazhou Ren, \IEEEmembership{Senior Member, IEEE}, Jianan Jiang, Chaoning Zhang, \IEEEmembership{Senior Member, IEEE}, Yang Yang, \IEEEmembership{Senior Member, IEEE}, and Heng Tao Shen, \IEEEmembership{Fellow, IEEE}
\thanks{This work was supported in part by National Natural Science Foundation of China (No. 62476052), Sichuan Science and Technology Program (Nos. 2024NSFSC1473 and 2024ZYD0268), and Shenzhen Science and Technology Program (Nos. JCYJ20230807120010021 and JCYJ20230807115959041). \textit{(Corresponding authors: Xiaorong Pu; Kecheng Chen.)}}
\thanks{Pengcheng Zheng and Chaoning Zhang are with the Center for Future Media and School
of Computer Science and Engineering, University of Electronic Science and Technology of China, Chengdu 611731, China.}
\thanks{Meng Yang, Bai Feng, and Jianan Jiang are with the Department of Computer Science and Engineering, University of Electronic Science and Technology of China, Chengdu 611731, China.}
\thanks{Yazhou Ren, and Xiaorong Pu are with the Department of Computer Science and Engineering, University of Electronic Science and Technology of China, Chengdu 611731, China and Shenzhen Institute for Advanced Study, University of Electronic Science and Technology of China, Shenzhen 518000, China (e-mail: puxiaor@uestc.edu.cn).}
\thanks{Kecheng Chen is with the Department of Electrical Engineering, and the Center for Intelligent Multidimensional Data Analysis, City University of Hong Kong, Hong Kong 999077, SAR (e-mail: cs.ckc96@gmail.com).}
\thanks{Jiaxin Huang is with the Department of Machine Learning, Mohamed bin
Zayed University of Artificial Intelligence, Abu Dhabi, United Arab Emirates.}
\thanks{Yang Yang is with the Center for Future Media, and the School of Computer Science and Engineering, University of Electronic Science and Technology of China, Chengdu 611731, China, and also with the Institute of Electronic and Information Engineering, University of Electronic Science and Technology of China, Guangdong 523808, China.}
\thanks{Heng Tao Shen is with the School of Computer Science and Technology, Tongji University, Shanghai 200092, China.}
}

\markboth{Journal of \LaTeX\ Class Files,~Vol.~14, No.~8, August~2021}%
{Shell \MakeLowercase{\textit{et al.}}: A Sample Article Using IEEEtran.cls for IEEE Journals}


\maketitle

\begin{abstract}
Recently, large language models (LLMs) have driven promising progress in lossless image compression. However, directly adopting existing paradigms for medical images suffers from an unsatisfactory trade-off between compression performance and efficiency. Moreover, existing LLM-based compressors often overlook the security of the compression process, which is critical in modern medical scenarios.
To this end, we propose a novel joint lossless compression and steganography framework. Inspired by bit plane slicing (BPS), we find it feasible to securely embed privacy messages into medical images in an invisible manner. Based on this insight, an adaptive modalities decomposition strategy is first devised to partition the entire image into two segments, providing global and local modalities for subsequent dual-path lossless compression. During this dual-path stage, we innovatively propose a segmented message steganography algorithm within the local modality path to ensure the security of the compression process. Coupled with the proposed anatomical priors-based low-rank adaptation (A-LoRA) fine-tuning strategy, extensive experimental results convincingly demonstrate the superiority of the proposed method in terms of compression ratios, efficiency, and security.
\end{abstract}


\section{Introduction}
\IEEEPARstart{I}{mage} compression task aims to reduce image size as much as possible, making it essential for high-quality data storage and transmission. Driven by the advancement of deep neural networks (DNNs), there has been remarkable progress in image compression, including lossy and lossless codecs \cite{tong2023learned}, \cite{zhang2024dnp}, \cite{liu2025learned}, \cite{liu2022multi}, \cite{Mao_2025_CVPR}, \cite{Lu_2025_CVPR}. Although existing lossy compression methods have shown remarkable performance for natural images, lossless compression for medical images has attracted massive attention from the medical imaging community, as even subtle degradations (\textit{e.g.,} mild blurring of tissues or organs) may compromise diagnostic accuracy \cite{wen2023msgfusion}, \cite{hou2025pathology}, \cite{Zhang2024ArIBBPS}, \cite{10478821}, \cite{li2025callic}, \cite{cai2024make}. 

\begin{figure}[!t]
 \includegraphics[width=\linewidth]{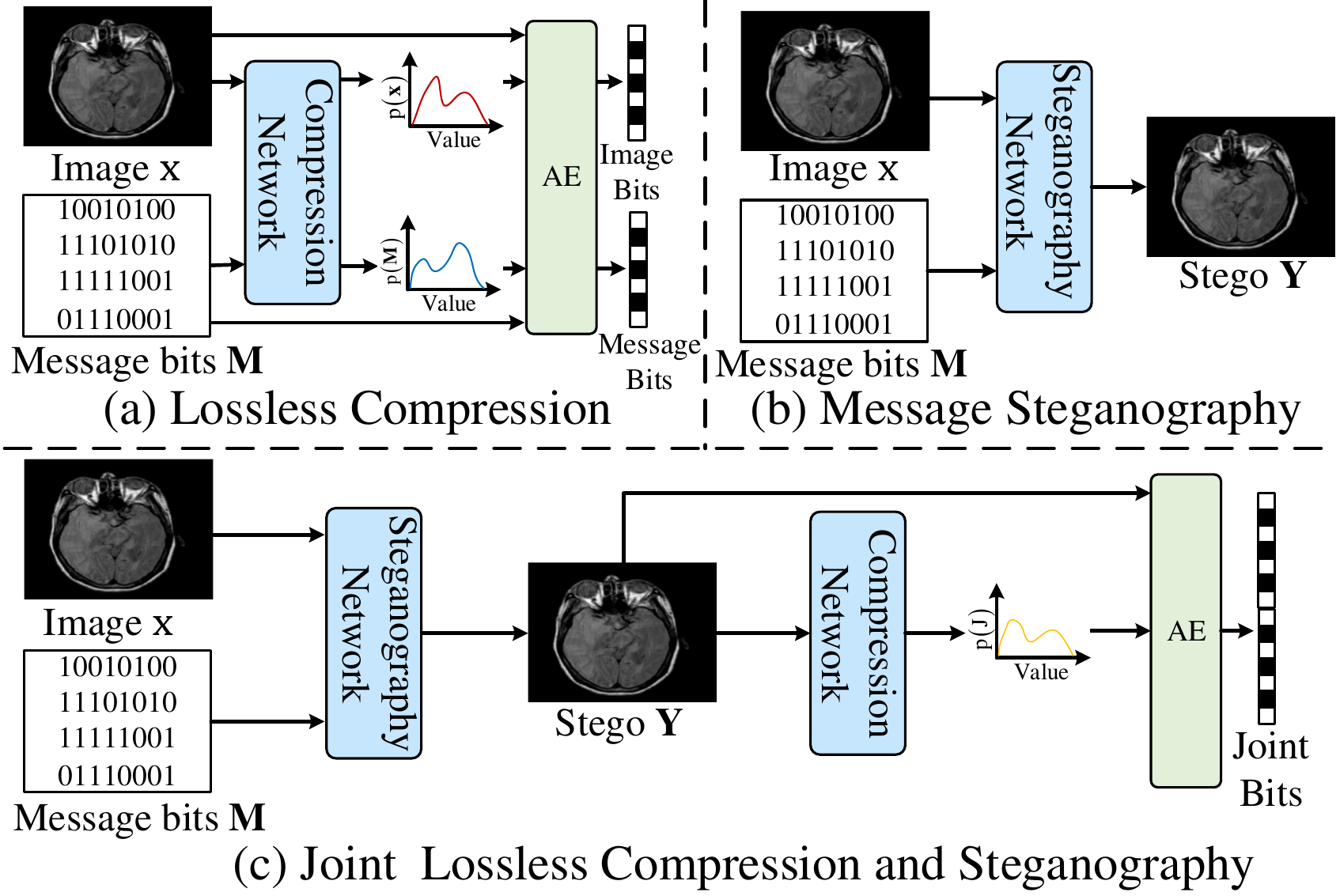}
\caption{Standard Lossless Compression and Steganography \textit{vs.} our proposed joint framework.}
\label{fig:intro}
\end{figure}

Recently, large language models (LLMs) have shown great success in probabilistic modeling and lossless compression~\cite{du2025largelanguagemodellossless}, \cite{chen2024largelanguagemodelslossless}. This is due to the so-called philosophy, \textit{“Intelligence” and “Compression" are two sides of the same coin} \cite{mackay2003information}. Theoretically, minimizing the negative $\log$-likelihood loss for next-token prediction is equivalent to maximizing the lossless compression objective \cite{heurtel2024compression}. Meanwhile, entropy coding in lossless compression seeks to accurately model data distribution to minimize the coding bitrate \cite{belyaev2013low}, \cite{Mao2025}, \cite{Valmeekam2023}, \cite{grangetto2006multimedia}, \cite{jakimoski2008cryptanalysis}. Consequently, LLMs could potentially provide an accurate probability distribution of images for entropy coding (\textit{e.g.,} arithmetic coding). Recently, this perspective is supported by Delétang et al. \cite{deletang2024language} who leverages the pre-trained LLMs to achieve competitive lossless compression ratios across various modalities, including text, audio and images. Whereas it can beat some domain-specific compressors by leveraging LLMs’ unprecedented intelligence, its encoding and decoding time are much slower than other baselines due to the inherent downside of autoregressive models. Compared with the autoregressive paradigm, latent variable models, \textit{e.g.,} variational autoencoder (VAE) often demonstrate impressive inference speed \cite{pinheiro2021variational}, \cite{child2021deepvaesgeneralizeautoregressive}. By utilizing the \textit{bits back} coding framework \cite{sugiyama2024initial} (predominantly, bits-back ANS), a VAE can be translated into a lossless codec. Whilst it provides an efficient lossless compression method, its compression ratio is always unsatisfactory \cite{Zhou_2018_CVPR_Workshops}, \cite{Jia_2024_CVPR}. 

With broad generality, autoregressive models (\textit{e.g.,} LLMs) can often be the most powerful but the slowest, while variational autoencoders are often the weakest but the fastest \cite{Tian2024, child2021deepvaesgeneralizeautoregressive}, \cite{An2024}, \cite{wang2019multi}, \cite{dai2009unified}. To this end, a potential option to address the above-mentioned issue is the combined use of autoregressive and latent variable models, achieving a great trade-off between compression performance and inference speed. Although existing combined methods \cite{Zhang2024ArIBBPS}, \cite{kang2022pilc} enable the efficient capture of both global and local data modalities, we argue that scaling them to the medical lossless image compression task is still unsatisfactory. First, these methods usually ignore the security of compressing medical images, where abundant meta information is attached to the images, \textit{e.g.}, patients' name, gender, ID number, and symptoms. For example, the DICOM standard \cite{varma2012managing} separately compresses image and its meta information, which exposes metadata in plaintext and makes it easier to intercept or tamper with. This approach puts patients' privacy at huge risk of being attacked by hackers when transmitted over the network. Second, most existing combined approaches implicitly learn global and local modalities from images, which makes it infeasible to embed the privacy message into the compression process, as even minimal perturbations of global modalities would change the overall appearance of the image. Third, existing combined methods never attempt to leverage the unprecedented intelligence capacity of LLMs, leading to a compromised compression ratio. \textit{Overall, existing methods struggle with maintaining a trade-off between sufficient compression (intelligence) performance and efficiency, and usually ignore the security for medical scenarios.}

To address the above question, we propose the joint lossless compression and steganography framework (as shown in Figure~\ref{fig:intro}) to ensure the security for compressing medical images, as well as rendering a great trade off between compression performance and efficiency. Unlike separate compression, which leaves metadata exposed as a parseable header~\cite{pei2014auxiliary}, our joint approach embeds it invisibly into the image bits. This removes any identifiable metadata segment, so an attacker can neither locate nor tamper with private information without first breaking the steganographic concealment. We empirically conduct a confirmatory experiment on bit plane slicing \cite{tang2016multiple} (BPS), which splits images in the bit plane dimension. As shown in Figure \ref{fig:bitplanes}, we find that the information amount in the image is gradually increasing from the least significant plane (LSP) to the most significant plane (MSP). Therefore, we pose an intuitive question: \textit{can we explicitly partition the image $\mathbf{x}$ into two segments, \textit{i.e.,} global and local modalities via BPS, thereby offering us a feasible way to embed the privacy message into the image in an invisible manner?} Obviously, the answer is affirmative. Specifically, we first devise the adaptive modalities decomposition strategy to partition the entire image using a slicing index, ranging in $\{1, ..., m\}$, denoted as $s$. To this end, the first segment comprises the local modalities, represented as $\mathbf{x}^{1: s}$, while the second segment includes the global modalities, \textit{e.g.,} low-frequency information, represented as $\mathbf{x}^{s+1: m}$, where $m$ is the bit-depth of image pixel. Second, considering that latent variable models cannot capture local features well, while a local autoregressive model is capable of doing so \cite{schirrmeister2020understanding}, \cite{shannon1948mathematical}, we propose the dual-path lossless compression scheme to improve the compression ratio. Notably, a novel segmented message steganography algorithm is designed in the local modality path to ensure a secure compression process for medical images. Finally, a simple yet effective fine-tuning strategy, \textit{i.e.,} anatomical priors-based low-rank adaptation (A-LoRA) is designed to further enhance the compression performance and accelerate the training speed. The embeddings of the global modality are used as visual prompts to fine-tune the LLM, which bridges the gap between the textual prior within LLM and visual understanding tasks.

Our main contributions can be summarized as follows:
\begin{itemize}
\item[$\bullet$] To explicitly extract the global and local modalities in medical images, we propose the adaptive modalities decomposition strategy, which lays a strong foundation for dual-path compression and provides a feasible way to embed the privacy message into the compression process.
\item[$\bullet$] A dual-path compression scheme is proposed to render a trade-off between sufficient compression performance and efficiency by leveraging LLM’s unprecedented intelligence (compression) capacity. Specifically, we devise a novel segmented message steganography algorithm in the local modality path, which embeds multiple message segments into their corresponding bit planes to ensure the security of the medical image compression process.
\item[$\bullet$] Moreover, an anatomical priors-based low-rank adaptation (A-LoRA) fine-tuning strategy is designed to enhance the compression performance and accelerate the training speed.
\item[$\bullet$] To the best of our knowledge, we are the first to explore the unprecedented intelligence of LLMs for medical images in a joint lossless compression and steganography framework. Extensive experimental results demonstrate the superiority of our proposed method in terms of compression performance, efficiency, and security.
\end{itemize}

\section{Related Work}
\subsection{Latent Variable Models for Compression} Latent variable models employ deep neural networks (DNNs) to map high-dimensional images into low-dimensional latent variables, integrating the \textit{bits-back} coding framework to render lossless image compression. For instance, L3C~\cite{mentzer2019practical} employs DNNs to extract latent variables as auxiliary information for lossless compression. However, the DNN-based transform is difficult to satisfy a tight bound on the maximum reconstruction error of each pixel. Therefore, Bai et al.~\cite{Bai_2021_SNIC} propose a new joint lossy image and residual compression framework for learning lossless compression. However, these methods are under the risk of posterior collapse. To this end, ArIB-BPS \cite{Zhang2024ArIBBPS} provides different importance for hierarchical latent variables, demonstrating superior compression performance with comparable inference speed. 

Nonetheless, latent variable models struggle to capture local features, resulting in suboptimal performance for images with intricate textures, \textit{e.g.,} medical images, where subtle details would significantly influence diagnostic accuracy.

\subsection{Large Language Models for Compression} LLMs have been widely adopted for computer vision tasks, \textit{e.g.,} image classification and segmentation~\cite{gou2024well}, \cite{yang2023improved} due to their accurate next token prediction capacity. Recently, literature~\cite{huang2024compression} implies the linear growth relation between compression performance and LLM’s intelligence, which provides a new perspective for lossless community. Theoretically, the negative $\log$-loss minimization for the next-token prediction of LLMs is equivalent to optimizing a lossless compression objective. Delétang et al. \cite{deletang2024language} are among the first to demonstrate that LLMs, when viewed as compressors, can outperform classical codecs like PNG~\cite{boutell1997png} in lossless compression for images, highlighting their strong potential in the lossless compression community. 

To the best of our knowledge, we are the first to leverage the unprecedented intelligence of LLMs in a joint lossless compression and steganography framework for medical images, achieving sufficient compression performance while maintaining security.
\vspace{-1mm}
\subsection{Generative Steganography}
Generative steganography has emerged as a promising direction for secure information hiding. Unlike conventional embedding-based steganography~\cite{chan2004hiding, holub2012designing}, which hides messages by modifying a cover, it conveys secret information through generated carriers, reducing explicit modification traces. Early studies focus on constructing image carriers from semantic structures, allowing secret information to be hidden during generation rather than directly embedded into a fixed cover image~\cite{zhou2023contour}. Subsequent works introduce reversible secret-to-image transformation, improving the recoverability and controllability of hidden messages in generated visual carriers~\cite{zhou2023s2irt}. Other studies explore text-based generative steganography, where long readable natural language sequences are generated as carriers for covert communication~\cite{cao2024longtext}. More recently, diffusion-based generative models have been adopted to improve the quality, robustness, and flexibility of stego carriers~\cite{hu2024gsdiffusion,li2024stegafds}.
\begin{figure*}[t]
 \centering
 \includegraphics[scale=0.35]{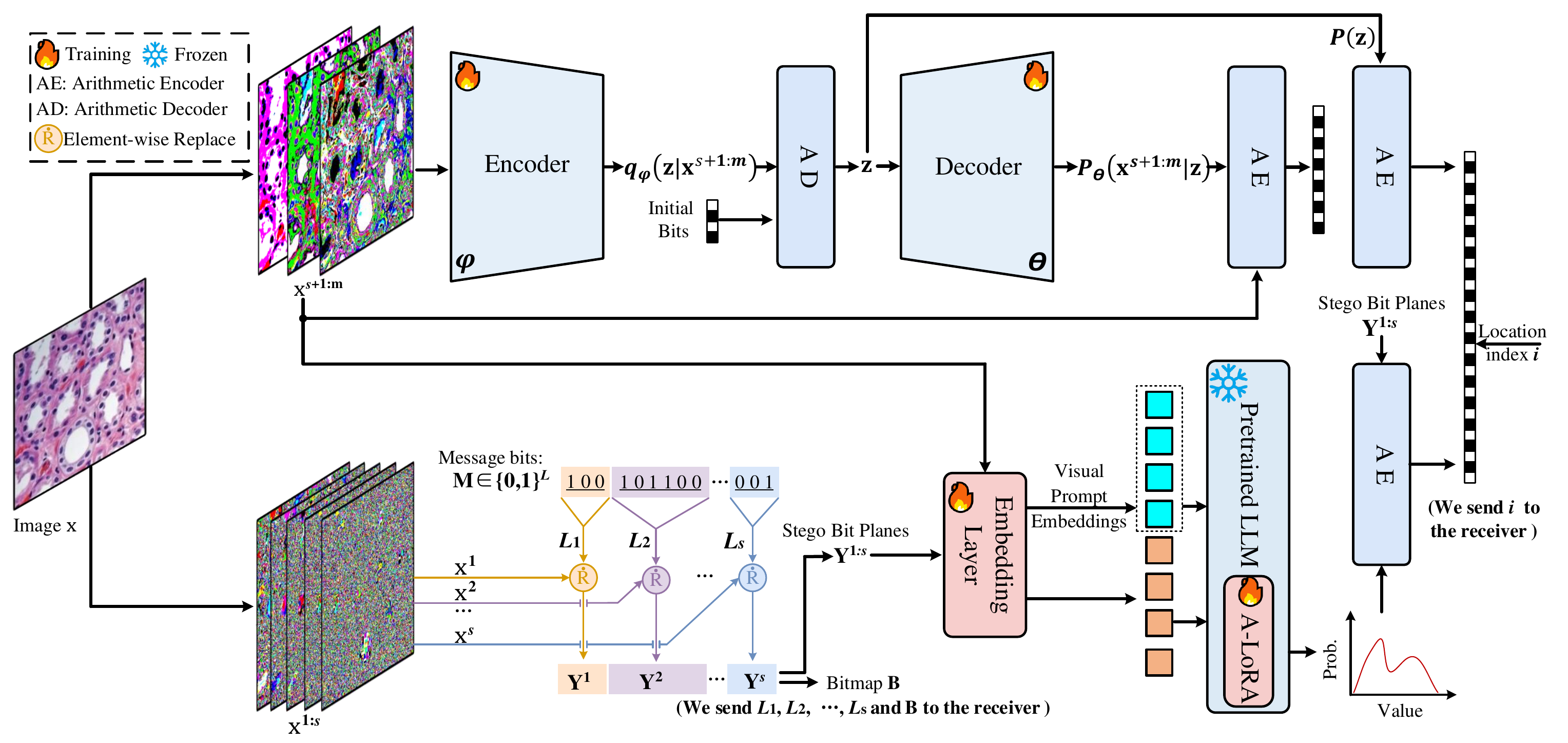}
 \caption{Overview of the proposed framework. In the encoding process, the image is firstly decomposed into global and local modalities via adaptive modalities decomposition strategy. Then, the global modalities are compressed using VAE with \textit{bits-back} coding scheme. The local modalities are compressed by leveraging the unprecedented intelligence of LLM. The final bitstream is concatenated from the dual-path outputs. The decoding order is the reverse of the encoding process.}
\label{fig:network}
\end{figure*}

These methods are related to our work in that both aim to improve the security of hidden communication. However, generative steganography mainly focuses on synthesized carriers, whereas our work targets medical image lossless compression with reversible message embedding. Different from those generation-based approaches, our framework emphasizes exact recovery of both the original medical image and the embedded message, while preserving diagnostically important image content during compression.


\section{Methodology}
\subsection{Preliminary and Overview}
\subsubsection{Preliminary} In this paper, we exploit two paradigms for lossless compression. Latent variable models approximate the true data distribution $p_{\text {data }}(\mathbf{x})$ with a marginal distribution $p_{\boldsymbol{\theta}}(\mathbf{x})$ defined by
\begin{equation}
\begin{aligned}  
p_{\boldsymbol{\theta}}(\mathbf{x}) = \int p_{\boldsymbol{\theta}}(\mathbf{x}, \mathbf{z}) \mathrm{d}\mathbf{z} = \int p_{\boldsymbol{\theta}}(\mathbf{x}|\mathbf{z}) p(\mathbf{z}) \mathrm{d}\mathbf{z},
\end{aligned}
\end{equation}
where $\mathbf{z}$ is an unobserved latent variable. Since exactly evaluating and optimizing $p_{\boldsymbol{\theta}}(\mathbf{x})$ is intractable, we can rewrite the marginal likelihood $p_{\boldsymbol{\theta}}(\mathbf{x})$ by introducing an \textit{inference model} $q_{\boldsymbol{\varphi}}(\mathbf{z} | \mathbf{x})$:
\begin{equation}
\begin{aligned}  
\log p_{\boldsymbol{\theta}}(\mathbf{x})=\underbrace{\mathbb{E}_{q_{\boldsymbol{\varphi}}(\mathbf{z}|\mathbf{x})} \log \frac{p_{\boldsymbol{\theta}}(\mathbf{x}, \mathbf{z})}{q_{\boldsymbol{\varphi}}(\mathbf{z} | \mathbf{x})}}_{\mathrm{ELBO}(q, \mathbf{x}; \boldsymbol{\theta}, \boldsymbol{\varphi })}+\underbrace{\mathbb{E}_{q_{\boldsymbol{\varphi}}(\mathbf{z} | \mathbf{x})} \log \frac{q_{\boldsymbol{\varphi}}(\mathbf{z}|\mathbf{x})}{p(\mathbf{z}| \mathbf{x})}}_{=D_{\mathrm{KL} }\left(q_{\boldsymbol{\varphi}}(\mathbf{z}|\mathbf{x}) \| p(\mathbf{z}|\mathbf{x})\right)}.
\end{aligned}
\end{equation}
As $D_{\mathrm{KL} }\left(q_{\boldsymbol{\varphi}}(\mathbf{z}|\mathbf{x}) \| p(\mathbf{z}|\mathbf{x})\right) \geq 0$, the inference and generative model can be found by jointly optimizing the Evidence Lower BOund (ELBO) (\textit{i.e.}, maximum $\mathrm{ELBO}(q, \mathbf{x}; \boldsymbol{\theta}, \boldsymbol{\varphi})$). With the help of the inference network $q_{\boldsymbol{\varphi}}(\mathbf{z} | \mathbf{x})$, it is possible to losslessly compress image $\mathbf{x}$ with \textit{bits back} coding via the following steps. \textbf{Step1}: Decode $\mathbf{z}$ with $q_{\boldsymbol{\varphi}}(\mathbf{z} | \mathbf{x})$ from initial bits. \textbf{Step2}: Encode $\mathbf{x}$ with $p_{\boldsymbol{\theta}}(\mathbf{x} | \mathbf{z})$. \textbf{Step3}: Encode $\mathbf{z}$ with $p(\mathbf{z})$. Proceeding in the reverse order, with the encode and decode operations swapped can losslessly recover the image. Another paradigm is autoregressive models. In this paper, we focus on LLMs, which minimize the negative $\log$-likelihood loss for next-token prediction. As a result, according to the Shannon’s source coding theorem \cite{yamamoto1994coding}, the cost of compressing the dataset $D=\left\{x_1, x_2, \ldots, x_n\right\}$ can be represented as:
\begin{equation}
\label{Eq3}
\begin{aligned}  
\left|S\right| &= -\log P_f(D) + \left|f\right| \\
&= -\sum_{i=1}^{n} \log p\left(x_i \mid x_1, x_2, \cdots, x_{i-1}\right) + \left|f\right| \\
&= \mathcal{L}_{LLM} + \left|f\right|,
\end{aligned}
\end{equation}
where $\left|f\right|$ is the code description of the compression method. Eq. \ref{Eq3} reveals that current LLM training protocols use a maximum-compression objective. It suggests that LLM could serve as a powerful tool for entropy coding, positioning the LLM as a general-purpose compressor for any modality.
\subsubsection{Overall Pipelines}
The overall framework of our proposed lossless image compression pipeline is illustrated in Figure~\ref{fig:network}. The original image $\mathbf{x}$ is firstly split into m-bit planes, adaptively producing the global modality (significant planes) $\mathbf{x}^{s+1: m}$ and local modality (insignificant planes) $\mathbf{x}^{1: s}$, where $s$ is the slicing index. Subsequently, we compress the global modality via variational autoencoders combined with \textit{bits back} coding scheme. For the local modality, LLM is employed to render an accurate probability representation of each encoded symbol for arithmetic coding. The embeddings of global modality are used as visual prompt for fine-tuning LLM, which bridges the gap between the textual prior within LLM and visual compression tasks. Meanwhile, we introduce a segmented message steganography algorithm when compressing the local modality. Finally, the bitstreams from the two paths, together with all required side information, are concatenated to form the final bitstream.

\subsection{Adaptive Modalities Decomposition}
\begin{figure}[t]
 \centering
 \includegraphics[width=\linewidth]{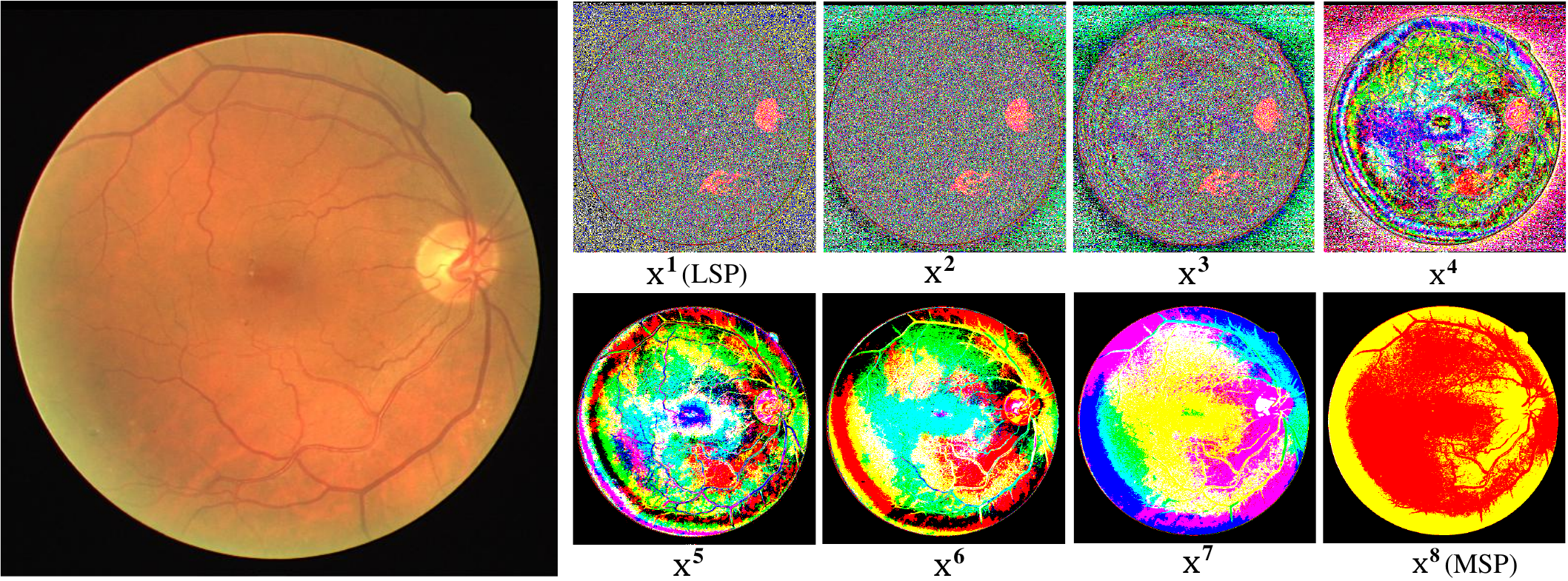}
 \caption{Visualize each bit plane, with the planes organized from the LSP to the MSP, arranged in a left-to-right and top-to-bottom manner.}
\label{fig:bitplanes}
\end{figure}

The current methods \cite{chen2024largelanguagemodelslossless}, \cite{deletang2024language} compress the entire image with LLM. Although these methods beat classical and learning image compression (LIC)-based codecs by leveraging the unprecedented intelligence within LLM. Their compression efficiency is unsatisfactory due to the inherent downside of autoregressive models. Moreover, they overlook security in medical image compression, where it is especially critical. To this end, it is urgent to achieve more efficient and secure compression for medical images. An intuitive idea is to partition the image into differently significant sub-parts. Specifically, given an image $\mathbf{x}$ with a bit depth of $m$, we can decompose it into $m$-binary bit planes via:
\begin{equation}
\begin{aligned}
\mathbf{x}^{l}=\left\lfloor \mathbf{x} / 2^{l-1}\right\rfloor \bmod 2,\ l\in\{1,\ldots,m\},
\label{decompose}
\end{aligned}
\end{equation}
\noindent where $\mathbf{x}^{l}$ is the $l^{th}$ bit plane, and $\lfloor \mathbf{x}\rfloor$ denotes the largest integer less than or equal to $\mathbf{x}$. The original image $\mathbf{x}$ can be reconstructed losslessly through the equation:
\begin{equation}
\begin{aligned}
\mathbf{x}=\sum_{l=1}^m 2^{l-1} \times \mathbf{x}^l,
\label{Recompose}
\end{aligned}
\end{equation}
\noindent Here, $\mathbf{x}^{1}$ denotes the LSP, while $\mathbf{x}^{m}$ denotes the MSP. As illustrated in Figure \ref{fig:bitplanes}, the volume of global data modality increases from the LSP to the MSP. Based on this finding, we adaptively partition the image into global modality $\mathbf{x}^{s+1:m}$ and local modality $\mathbf{x}^{1:s}$ with a slicing index $s$, which can be determined by:
\begin{equation}
\begin{aligned}
s^*=\sup \left\{ s \,\middle|\, \sum_{i=1}^s I\left(\mathbf{x}^i; \mathbf{x}\right) \leq (1-\beta) * I\left(\mathbf{x}; \mathbf{x}\right) \right\},
\label{slicing index}
\end{aligned}
\end{equation}
\noindent where $I(\mathbf{x}^i; \mathbf{x})$ denotes the mutual information between the $i$-th bit plane and the original image, and $I(\mathbf{x}; \mathbf{x})$ equals the entropy of the image $\mathbf{x}$, \textit{i.e.}, the total information content of the image. $\beta \in [0,1]$ is a hyperparameter controlling information retention, which is ablated in Section IV-C2. $\sup$ denotes the upper bound, ensuring that the largest slicing index $s^*$ is selected such that the cumulative mutual information does not exceed the target information value. In practice, the mutual information term in Eq.~\ref{slicing index} is estimated using empirical discrete distributions at the pixel level. Specifically, for the $i$-th bit plane $\mathbf{x}^{i}$, we collect paired samples $(\mathbf{x}^{i}(p), \mathbf{x}(p))$ over all pixel locations $p$ in the training set, where $\mathbf{x}^{i}(p)\in\{0,1\}$ denotes the binary value of the $i$-th bit plane at pixel $p$, and $\mathbf{x}(p)\in\mathcal{X}$ denotes the corresponding original image intensity. Based on these paired samples, we construct the empirical joint distribution $\hat{p}(\mathbf{x}^{i}, \mathbf{x})$ and the corresponding marginals $\hat{p}(\mathbf{x}^{i})$ and $\hat{p}(\mathbf{x})$, and compute
\begin{equation}
\small
\begin{aligned} 
I(\mathbf{x}^{i}; \mathbf{x})=\sum_{a\in\{0,1\}}\sum_{b\in\mathcal{X}} \hat{p}(\mathbf{x}^{i}=a, \mathbf{x}=b)\log\frac{\hat{p}(\mathbf{x}^{i}=a, \mathbf{x}=b)}{\hat{p}(\mathbf{x}^{i}=a)\hat{p}(\mathbf{x}=b)+\epsilon},
\end{aligned}
\end{equation}
where $a$ and $b$ enumerate the possible values of $\mathbf{x}^{i}$ and $\mathbf{x}$, respectively, $\mathcal{X}$ denotes the discrete image intensity space, and $\epsilon$ is a small constant for numerical stability, set to $10^{-12}$ in our implementation. After estimating $I(\mathbf{x}^{i};\mathbf{x})$ for each bit plane, we accumulate these values from the LSP to the MSP and select the largest slicing index $s^*$ that satisfies the information-retention constraint in Eq.~\ref{slicing index}.

\noindent\textit{\textbf{Discussion}} - \textit{Why $I(\mathbf{x}; \mathbf{x})$ is equal to $H(\mathbf{x})$, \textit{i.e.}, the total information content of the image.}
In information theory, the mutual information $I(\mathbf{x}; \mathbf{y})$ between two discrete random variables $\mathbf{x}$ and $\mathbf{y}$ is defined as:
\begin{equation}
I(\mathbf{x}; \mathbf{y}) = \sum_{x \in \mathcal{X}} \sum_{y \in \mathcal{Y}} p(x,y) \log \frac{p(x,y)}{p(x)p(y)}.
\end{equation}
When $\mathbf{y} = \mathbf{x}$, the joint distribution $p(x,x)$ degrades to $p(x)$, and the conditional probability $p(x|x) = 1$. Substituting these into the mutual information definition:
\begin{align}
I(\mathbf{x}; \mathbf{x}) &= \sum_{x \in \mathcal{X}} p(x) \log \frac{p(x)}{p(x)p(x)} \notag \\
&= \sum_{x \in \mathcal{X}} p(x) \log \frac{1}{p(x)} \notag \\
&= H(\mathbf{x}),
\end{align}
where $H(\mathbf{x}) = -\sum_{x \in \mathcal{X}} p(x) \log p(x)$ is the entropy of $\mathbf{x}$. This equality holds because the self-information $\log \frac{1}{p(x)}$ exactly corresponds to the contribution of each outcome to the entropy. Therefore, the mutual information between a random variable and itself equals its entropy, representing the total information content. This fundamental property is utilized in the paper's adaptive modalities decomposition to quantify and partition the image information. Moreover, we provide pseudocode, Algorithm \ref{alg:bitplane} and Algorithm \ref{alg: adaptivebitplane} for each method bit plane slicing algorithm and adaptive modalities decomposition algorithm, respectively.

\begin{algorithm}[h]
\caption{Bit Plane Slicing Algorithm}
\begin{algorithmic}[1]
\footnotesize
    \State \textbf{Input:} \texttt{image} $\mathbf{x}$ (tensor), \texttt{bits} $m$ (integer)
    \State \textbf{Output:} \texttt{bit\_planes} = [$\mathbf{x}^{1}$, $\mathbf{x}^{2}$, ..., $\mathbf{x}^{m}$] (ordered from LSP to MSP) 

    \Function{bit\_slicing}{$\mathbf{x}$, $m$}
        \State \texttt{bit\_planes} $\gets$ [ ] \Comment{Initialize an empty list}
        
        \For{$i = 1$ to $m$}
            \State \texttt{bit\_planes.append($\mathbf{x}$ \% 2)}
            \State $\mathbf{x}$ $\gets$ $\mathbf{x}$ // 2 \Comment{Integer division by 2}
        \EndFor
        \State \Return \texttt{bit\_planes} \Comment{Return bit-planes}
    \EndFunction
    \end{algorithmic}
    \label{alg:bitplane}
\end{algorithm}

\begin{algorithm}[!t]
\caption{Adaptive Modalities Decomposition Algorithm}
\begin{algorithmic}[1]
\footnotesize
    \State \textbf{Input:} \texttt{image} $\mathbf{x}$ (tensor), $\beta$ (float), mutual information score function \texttt{I($\cdot$)}, \texttt{bits} $m$ (integer)
    \State \textbf{Output:} local modality $\mathbf{x}^{1: s^*}$, global modality $\mathbf{x}^{s^*+1: m}$
    
    \State Initialize: cumulative mutual information amount: $T_{cu} \gets 0$, whole information amount in the image: $T_{w} \gets \texttt{I}(\mathbf{x}; \mathbf{x})$, target retentive information amount: $T_{t} \gets  (1-\beta)*T_{w}$, slicing index: $s^* \leftarrow 0$
    \For{$s = 1$ to $m$}
        \State Calculate the information amount in the current bit
plane: $T_{c} \gets \texttt{I}(\mathbf{x}^{s}; \mathbf{x})$
        \State \textbf{if} {$T_{cu}+T_{c} \leq T_{t}$} \textbf{then} $T_{cu} \gets T_{cu}+T_{c}$, $s^* \leftarrow s$
        \State \textbf{else}\ break
        \State \textbf{end if}
    \EndFor
    \State \Return {$\mathbf{x}^{1: s^*}$, $\mathbf{x}^{s^*+1: m}$}
    \end{algorithmic}
    \label{alg: adaptivebitplane}
\end{algorithm}





\subsection{Dual-Path Lossless Compression}
Among all current algorithms, latent variable models are often the weakest but the fastest, while LLM-based models can often be the most powerful but the slowest. As suggested by many previous works \cite{schirrmeister2020understanding, shannon1948mathematical}, latent variable models cannot capture local modalities, while LLM-based models are capable of doing so. Thus, an intuitive idea would be to exploit LLM to compress local modalities, and a latent variable model for more global ones. This design can not only achieve a great trade off between model efficiency and performance but also allows us to embed the privacy message into the image in an invisible manner.

\subsubsection{Latent Variable Path for Global Modality}
By recalling the paradigm of latent variable model with \textit{bits back} coding scheme in the preliminary, we compress the global modality via the following steps. \textbf{Step1}: Decode $\mathbf{z}$ with $q_{\boldsymbol{\varphi}}(\mathbf{z} | \mathbf x^{s+1: m})$ from initial bits, where initial bits are generated from the auxiliary random source. \textbf{Step2}: Encode $ \mathbf x^{s+1: m}$ with $p_{\boldsymbol{\theta}}(\mathbf x^{s+1: m} | \mathbf{z})$. \textbf{Step3}: Encode $\mathbf{z}$ with $p(\mathbf{z})$. The posterior distribution $q_{\boldsymbol{\varphi}}(\mathbf{z} | \mathbf x^{s+1: m})$ and the likelihood probability distribution $p_{\boldsymbol{\theta}}(\mathbf x^{s+1: m} | \mathbf{z})$ are predicted by the residual block based encoder parameterized by $\varphi$ and the residual block based decoder parameterized by $\theta$, respectively. The decompression order is the reverse of the compression process, \textit{i.e.,} \textbf{Step1}: decode $\mathbf{z}$ with $p(\mathbf{z})$. \textbf{Step2}: decode $ \mathbf x^{s+1: m}$ with $p_{\boldsymbol{\theta}}(\mathbf x^{s+1: m} | \mathbf{z})$. \textbf{Step3}: Encode $\mathbf{z}$ with $q_{\boldsymbol{\varphi}}(\mathbf{z} | \mathbf x^{s+1: m})$.

\subsubsection{LLM Path for Local Modality}
\begin{figure}[t]
 \centering
 \includegraphics[width=\linewidth]{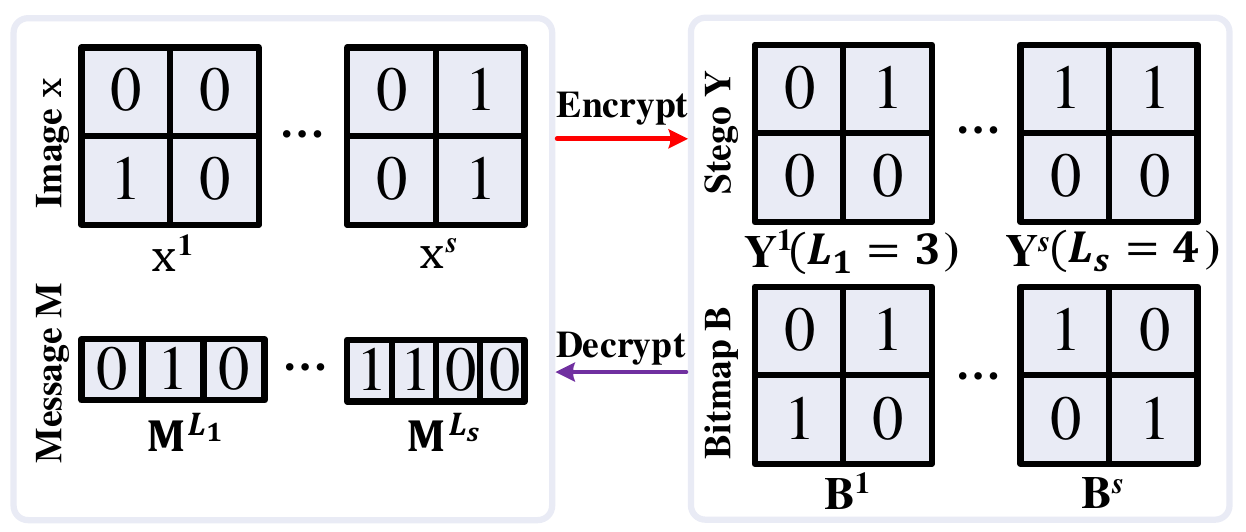}
 \caption{An example of the encryption and decryption process. We send $L_1$, $L_2$, $L_s$, and Bitmap $\mathbf{B}$ to the receiver to ensure image $\mathbf{x}$ and message $\mathbf{M}$ can be losslessly recovered.}
\label{fig:encryption}

\end{figure}

Our work is not blind steganography for arbitrary public communication scenarios, but rather a joint lossless compression and reversible message embedding framework for medical images, where the exact recovery of both the medical image and the embedded message is required by authorized receivers. Under this application setting, non-blind extraction is a deliberate design choice to guarantee strict reversibility, rather than a limitation overlooked by the method. In practical medical-image transmission and archiving scenarios, the receiver is typically an authorized party that already performs full decoding of the transmitted bitstream, consistent with standard medical-imaging workflows and access-control requirements in DICOM-based systems and healthcare data protection regulations~\cite{varma2012managing}, \cite{assistance2003summary}. Therefore, the required side information is compatible with the target deployment scenario, rather than an impractical overhead. In other words, our framework prioritizes lossless recovery and compression compatibility, instead of pursuing fully blind extraction at the cost of reversibility.
Since the information of local modality is relatively insignificant, modifying it will generally not affect the overall appearance of the image. To achieve an invisible and more secure message steganography, we carefully design a segmented message steganography algorithm in this path, which embeds the message into the medical image. Given the binary identity message $\mathbf{M} \in\{\mathbf{0}, \mathbf{1}\}^{L}$ with a sampling length of $L$, we first split it into $s$ sub-message segments. The length of $i$-th message segments is represented as $L_i$, where $\sum_{i=1}^s L_i=L$. Rather than the vanilla LSB steganography algorithm \cite{neeta2006implementation}, which directly replaces each pixel value in the LSP (\textit{i.e.,} the 1-th bit planes) with the bits from the entire secret message, we replace each binary pixel value of $\mathbf{x}^{i}$ with its corresponding message segment $\mathbf{M}^{L_i}$. Figure \ref{fig:encryption} gives an example of the encryption and decryption process. The encryption process of our proposed segmented message steganography algorithm can be formulated by
\begin{equation}
\begin{aligned}  
\{\mathbf{Y}^i\}_{i=1}^s &= \{\mathbf{x}^i\}_{i=1}^s \,\, \mathbf{\dot{R}}_{\{L_{i}\}_{i=1}^s} \,\, \{\mathbf{M}^{Li}\}_{i=1}^s, \\
\{\mathbf{B}^i\}_{i=1}^s &= \{\mathbf{x}^i\}_{i=1}^s \oplus \{\mathbf{Y}^i\}_{i=1}^s, 
\end{aligned}
\label{en}
\end{equation}
where $\mathbf{Y}^i$ and $\mathbf{B}^i$ represent the $i$-th stego bit plane and the $i$-th bitmap, respectively. We define $\mathbf{\dot{R}}_{L_{i}}$ as a novel operation, which uses the $L_{i}$-length binary message of the segmented message $\mathbf{M}^{Li}$ to replace the binary value of the $i$-th bit plane $\mathbf{x}^i$ element by element. When performing the replace operation, we follow the sequence from top to bottom, and within each row, from left to right. The symbol $\oplus$ is the XOR operation. With the aid of $\{L_{i}\}_{i=1}^s$ and $\{\mathbf{B}^i\}_{i=1}^s$, we can recover the original message $\{\mathbf{M}^{Li}\}_{i=1}^s$ and image $\{\mathbf{x}^i\}_{i=1}^s$, respectively:
\begin{equation}
\begin{aligned}  
\{\mathbf{M}^{Li}\}_{i=1}^s &= \{\mathbf{Y}^i\}_{i=1}^s \,\,  \mathbf{\dot{E}}_{\{L_{i}\}_{i=1}^s}, \\
\{\mathbf{x}^i\}_{i=1}^s &= \{\mathbf{Y}^i\}_{i=1}^s \oplus \{\mathbf{B}^i\}_{i=1}^s,
\end{aligned}
\label{de}
\end{equation}
where $\mathbf{\dot{E}}_{L_{i}}$ represents extracting $L_{i}$-length binary message from $\mathbf{Y}^i$. The extract sequence is also from top to bottom, and within each row, from left to right. 

Although our proposed method does not explicitly optimize a texture- or semantics-aware embedding cost as in dedicated steganographic frameworks, it does not ignore image significance. Specifically, the message is embedded only into the low-significance local modality rather than the diagnostically dominant global modality through adaptive bit-plane decomposition. Therefore, our proposed method already incorporates image significance information into the embedding process in a manner that is consistent with the reversibility and lossless compression requirements. Unlike vanilla LSB steganography, which directly replaces the pixel values in the LSP with the entire secret message, our proposed method first decomposes the image into global and local modalities, and then splits the secret message into multiple sub-message segments that are embedded into several local bit planes. This design reduces concentrated perturbation on any single bit plane and preserves the high-information global modality, which is theoretically more favorable for concealment. The corresponding formulation is given in Eq.~(\ref{en})-(\ref{de}), and the encryption/decryption process is illustrated in Figure~\ref{fig:encryption}. The bitmap $\mathbf{B}$ in Eq.~(\ref{en})-(\ref{de}) is not a fixed or shared pattern across all cover images. Instead, it is an image-specific auxiliary recovery map, computed from the XOR operation between the original local bit planes and the corresponding stego bit planes, \textit{i.e.,}
$\{\mathbf{B}^i\}_{i=1}^s = \{\mathbf{x}^i\}_{i=1}^s \oplus \{\mathbf{Y}^i\}_{i=1}^s$. Therefore, $\mathbf{B}$ is determined by the specific image content and the embedded message, and varies from one cover image to another.

After obtaining the stego bit planes, a learnable embedding layer with 4-convolution layers is employed to extract image embeddings. For the image compression task, the embeddings of global modality are used as visual prompt, which supplies the LLM with visual information about the image. This design allows the model to integrate image information with the LLM’s prior knowledge, bridging the gap between image and text tasks, ultimately enhancing compression efficiency. Following \cite{chen2024largelanguagemodelslossless}, we leverage the two-step lossless pixel tokenization strategy for pixel sequence semantic understanding, and the predictive distribution sampling algorithm is employed to obtain the probability distribution of each pixel from the output of LLM. With the encoded pixel values and corresponding probability distribution, we can conduct arithmetic encoder to get the bitstream. 
Specifically, after segmented message steganography, the local modality is converted from $\mathbf{x}^{1:s}$ to the stego local modality $\mathbf{Y}^{1:s}=\{\mathbf{Y}^1,\mathbf{Y}^2,\ldots,\mathbf{Y}^s\}$, where each $\mathbf{Y}^j \in \{0,1\}^{H\times W}$ is a binary bit plane. Since the local modality is fed into the LLM in a patch-wise manner, we first partition each stego bit plane into $K$ non-overlapping patches of size $P\times P$, where $P$ denotes the patch size. Let $\mathbf{Y}_k^j \in \{0,1\}^{P\times P}$ denote the $k$-th patch of the $j$-th stego bit plane, where $k\in\{1,2,\ldots,K\}$. To clearly define the input to the LLM-based compression path, we serialize each local patch into a one-dimensional binary sequence in a fixed raster-scan order. Specifically, each local patch is first flattened row by row into a binary sequence as
\begin{equation}
\begin{aligned}
\mathrm{vec}(\mathbf{Y}_k^j)=[y_{k,1}^j,y_{k,2}^j,\ldots,y_{k,P^2}^j], \quad y_{k,t}^j\in\{0,1\},
\end{aligned}
\end{equation}
where $\mathbf{Y}_k^j \in \{0,1\}^{P\times P}$ denotes the $k$-th patch of the $j$-th stego bit plane, $y_{k,t}^j$ denotes the $t$-th binary symbol in the raster-scanned sequence of $\mathbf{Y}_k^j$, and $t\in\{1,2,\ldots,P^2\}$. The binary sequence of the $k$-th local patch is then constructed by concatenating all of its stego bit planes from the first to the $s$-th plane:
\begin{equation}
\begin{aligned}
\mathcal{S}_{\mathrm{bit}}^{(k)}=\mathrm{vec}(\mathbf{Y}_k^1)\,\Vert\, \mathrm{vec}(\mathbf{Y}_k^2)\,\Vert\, \cdots \,\Vert\, \mathrm{vec}(\mathbf{Y}_k^s),
\end{aligned}
\end{equation}
where $\,\Vert\,$ denotes concatenation. Therefore, the LLM compresses the local modality as a patch-wise autoregressive binary-symbol prediction problem. To define lossless tokens, each binary symbol in $\mathcal{S}_{\mathrm{bit}}^{(k)}$ is treated as a textual token with a one-to-one mapping. Let $\mathcal{T}(\cdot)$ denote the tokenizer of the pretrained LLM and $\mathcal{D}$ its vocabulary. We define a binary token dictionary
\begin{equation}
\begin{aligned}
\mathcal{D}_{\mathrm{bin}}=\{\mathcal{T}(\texttt{"0"}),\mathcal{T}(\texttt{"1"})\},
\end{aligned}
\end{equation}
so that the two possible bit values are uniquely mapped to two language tokens. This preserves exact reversibility and avoids ambiguity in symbol-token conversion. For context modeling, we use an autoregressive next-symbol prediction scheme. Let $\mathcal{P}$ denote a fixed task prompt describing that the input is a flattened binary sequence from stego bit-plane patches and that the model should predict the next binary symbol conditioned on previous ones. For the $t$-th symbol in $\mathcal{S}_{\mathrm{bit}}^{(k)}$, the LLM input context is constructed as $\mathcal{C}_t^{(k)}=\mathcal{P}\,\Vert\,(\mathcal{S}_{\mathrm{bit}}^{(k)})^{<t},$
where $(\mathcal{S}_{\mathrm{bit}}^{(k)})^{<t}$ denotes the prefix of all previously encoded binary symbols within the current patch sequence. In addition, the embeddings of the global modality are injected as visual prompt embeddings into the embedding layer, providing anatomical and structural priors to guide the prediction of local binary symbols. Given the context $\mathcal{C}_t^{(k)}$, the LLM outputs logits over the whole vocabulary. We then extract only the two logits corresponding to $\texttt{"0"}$ and $\texttt{"1"}$ from $\mathcal{D}_{\mathrm{bin}}$, and normalize them into a binary probability distribution:
\begin{equation}
\begin{aligned}
p_t^{(k)}(v)=\frac{\exp(\ell_t^{(k)}(v))}{\exp(\ell_t^{(k)}(0))+\exp(\ell_t^{(k)}(1))}, \quad v\in\{0,1\},
\end{aligned}
\end{equation}
where $\ell_t^{(k)}(v)$ is the logit of symbol $v$ at step $t$ for the $k$-th patch sequence. The resulting binary distribution $p_t^{(k)}$ is finally used by the arithmetic encoder to encode the current bit. During decompression, the same prompt, tokenizer, visual prompt embeddings, and autoregressive order are used to reproduce the probability distribution and decode each patch sequence losslessly.
In practical deployment, $\mathbf{B}$, together with $\{L_i\}$ and the location index $i$, is appended to the final compressed bitstream for authorized decoding, and can be further protected by standard encryption or secure encapsulation mechanisms (\textit{e.g.,} TLS 1.3~\cite{rescorla2018transport} or IPsec~\cite{kent2005security}) if required by the application scenario. Therefore, the security of the framework does not rely on exposing $\mathbf{B}$ as a standalone public bitmap. More specifically, the role of the proposed steganographic design is to avoid leaving the private message as an explicit and easily parseable component during the joint compression process, whereas confidentiality against interception can be further enforced by standard secure-transmission mechanisms when necessary in practical deployment.

\subsubsection{Discussion - Why Segmented Message Steganography Improves Security}
To further provide theoretical support for the security enhancement of the proposed segmented message steganography, we analyze the embedding perturbation from the perspectives of modification concentration, distortion upper bound, and statistical detectability.
Let $\mathbf{x}_i(p)$ and $\mathbf{Y}_i(p)$ denote the bit values at pixel location $p$ in the $i$-th local bit plane before and after embedding, respectively, where $i \in \{1,2,\ldots,s\}$. Here, $s$ denotes the number of local bit planes used for message embedding, and $p$ indexes the pixel position in an image of spatial size $H \times W$, where $H$ and $W$ are the image height and width, respectively. We define the modification flag on the $i$-th bit plane at position $p$ as
\begin{equation}
\Delta_i(p)=
\begin{cases}
1, & \text{if } \mathbf{x}_i(p)\neq \mathbf{Y}_i(p),\\
0, & \text{otherwise}.
\end{cases}
\end{equation}
Here, $\Delta_i(p)$ indicates whether the bit at location $p$ in the $i$-th bit plane is changed by message embedding.
Accordingly, the average modification rate on the $i$-th bit plane is given by
\begin{equation}
r_i = \sum_{p=1}^{H\times W} \Delta_i(p) / (H \times W).
\end{equation}
Here, $r_i$ denotes the proportion of modified bits on the $i$-th local bit plane, and $H \times W$ is the total number of pixels in the image.
For a vanilla single-plane replacement strategy, the entire payload is concentrated on one LSP, which leads to a relatively large peak modification rate on that plane. In contrast, our segmented strategy splits the secret message into multiple sub-message segments and distributes them across several low-significance local bit planes. Let $\{L_i\}_{i=1}^{s}$ denote the lengths of the message segments embedded into different bit planes, where $L_i$ is the number of embedded message bits assigned to the $i$-th bit plane and $\sum_{i=1}^{s} L_i = L$, with $L$ being the total message length. Under the same total payload, the segmented strategy reduces the perturbation concentration on any individual bit plane, \textit{i.e.,} it yields a smaller peak modification rate $\max_i r_i$ than direct single-plane replacement. This property is beneficial because concentrated perturbations are more likely to introduce detectable statistical artifacts~\cite{pevny2010using},~\cite{holub2014universal},~\cite{fridrich2012rich}.
Moreover, since the embedding is restricted to the low-significance local modality, the resulting pixel-level distortion is explicitly bounded by the contribution of the low-order bit planes only. Specifically, the distortion at pixel location $p$ can be expressed as
\begin{equation}
D(p) = \sum_{i=1}^{s} 2^{i-1}\Delta_i(p).
\end{equation}
Here, $D(p)$ denotes the accumulated distortion at pixel $p$, and $2^{i-1}$ is the value contribution of the $i$-th bit plane to the pixel intensity. This indicates that the diagnostically dominant high-significance bit planes remain unaffected. This significance-aware constraint reduces the impact of embedding on the overall image statistics and visual structure.
From a statistical perspective, let $P_c$ and $P_s$ denote the feature distributions of the cover and stego images, respectively. A more secure embedding strategy should induce a smaller distribution shift between $P_c$ and $P_s$. Compared with direct single-plane replacement, the proposed segmented embedding introduces weaker concentrated perturbations and preserves the dominant global modality, thereby leading to a smaller cover--stego divergence and lower detectability by steganalysis models under matched payload settings.

\subsection{Anatomical Priors-based Fine-Tuning}
The naive LoRA fine-tuning \cite{hu2022lora} algorithm uses a random Gaussian initialization for $A$ and zero for $B$, which does not take into account the inherent priors (\textit{e.g.,} the statistical properties of anatomical features) within medical images. To further enhance the compression performance and accelerate the fine-tuning process, we design a simple yet effective fine-tuning strategy, coined as anatomical priors-based fine-tuning (A-LoRA). In A-LoRA, we first use the pretrained MobileNetV3-S \cite{howard2019searching} to extract the anatomical features from medical image dataset. Then, the mean $\mu_{\mathrm{f}}$ and var $\sigma_{\mathrm{f}}^2$ of anatomical features are leveraged to initialize the $A$, \textit{i.e.}, $A=\mathcal{N}\left(\mathbf{Y}^{1: s}; \mu_{\mathrm{f}}, \sigma_{\mathrm{f}}^2\right)$. This simple design guides the adaptation process of LLM in the field of medical image processing, resulting in a faster fine-tuning speed and better compression performance.
\begin{table}[!t]
\caption{Datasets introduction. $*$ and $\dagger$ indicate training and testing datasets, respectively. C-19-R and C-19-C refer to the COVID-
19 radiograph dataset, and COVID-19 CT scan dataset, respectively. L-Spine is the Lumbar Spine dataset. B-Ultra refers to the Breast Ultrasound dataset.}
\centering
\small
\begin{tabular}{@{}cccc@{}}
\toprule
Dataset & Description &\# Num. & Avg. Resolution \\
\midrule
Flickr2k*                      & Natural         & 2650  & 2048$\times$1024 \\
C-19-R$\dagger$                & Chest X-ray     & 42300 & 300$\times$300   \\
C-19-C$\dagger$                & Lung CT         & 262   & 512$\times$512   \\
Lung CT$\dagger$               & Lung CT         & 7944  & 1024$\times$768  \\
Head CT$\dagger$               & Head CT         & 200   & 640$\times$480   \\
L-Spine$\dagger$          & Spine CT        & 2477  & 256$\times$256   \\
B-Ultra$\dagger$     & Ultrasound      & 1578  & 640$\times$480   \\
Brain Tumor$\dagger$           & Brain MRI       & 7022  & 512$\times$512   \\
\bottomrule
\end{tabular}
\label{tb:dataset_details}
\end{table}
\begin{table*}[!t]
    \small
    \setlength{\tabcolsep}{6pt}
    \centering
    \caption{Compression performance of the proposed methods and other codecs in terms of bits per pixel (bpp). Notably, the reported bpp for our method includes all side information required for exact reconstruction, including $\{L_{i}\}$, $\mathbf{B}$ and $i$. We show the difference in percentage to our approach, using green. The best is highlighted in bold, and the second is underlined.}
    \definecolor{lightgray}{RGB}{240,240,240}
    \definecolor{gaingreen}{RGB}{0,150,0}
    
    \newcommand{\graycell}[1]{\cellcolor{lightgray}#1}
    
    \begin{tabular}{clllllll}
    \toprule
    Codec & {C-19-R} & {C-19-C} & {Lung CT} & {Head CT} & {L- Spine} & {B-Ultra} & {Brain Tumor} \\
    \midrule
    PNG (1997)             & 3.10 \textcolor{gaingreen}{\scriptsize +37.8\%} & 4.56 \textcolor{gaingreen}{\scriptsize +9.1\%} & 4.32 \textcolor{gaingreen}{\scriptsize +136.1\%} & 3.70 \textcolor{gaingreen}{\scriptsize +140.3\%} & 3.53 \textcolor{gaingreen}{\scriptsize +35.3\%} & 3.45 \textcolor{gaingreen}{\scriptsize +29.2\%} & 3.52 \textcolor{gaingreen}{\scriptsize +29.4\%} \\
    JPEG-LS (TIP'2000)             & 2.75 \textcolor{gaingreen}{\scriptsize +22.2\% } & 4.41 \textcolor{gaingreen}{\scriptsize +5.5\% } & 4.21 \textcolor{gaingreen}{\scriptsize +130.1\%} & 4.51 \textcolor{gaingreen}{\scriptsize +192.9\%} & 3.89 \textcolor{gaingreen}{\scriptsize +47.4\% } & 3.05 \textcolor{gaingreen}{\scriptsize +14.2\%} & 3.31 \textcolor{gaingreen}{\scriptsize +21.7\%} \\
    JPEG2000 (2000)        & 2.77 \textcolor{gaingreen}{\scriptsize +23.1\%} & 4.47 \textcolor{gaingreen}{\scriptsize +6.9\%} & 4.52 \textcolor{gaingreen}{\scriptsize +147.0\%} & 4.66 \textcolor{gaingreen}{\scriptsize +202.6\%} & 4.16 \textcolor{gaingreen}{\scriptsize +59.4\%} & 3.01 \textcolor{gaingreen}{\scriptsize +12.7\%} & 3.34 \textcolor{gaingreen}{\scriptsize +22.8\%} \\
    WebP (2015)            & 2.79 \textcolor{gaingreen}{\scriptsize +24.0\%} & 4.35 \textcolor{gaingreen}{\scriptsize +4.1\%} & 3.52 \textcolor{gaingreen}{\scriptsize +92.4\%} & 3.00 \textcolor{gaingreen}{\scriptsize +94.8\%} & 3.30 \textcolor{gaingreen}{\scriptsize +26.4\%} & 3.08 \textcolor{gaingreen}{\scriptsize +15.4\%} & 3.27 \textcolor{gaingreen}{\scriptsize +20.2\%} \\
    FLIF (ICIP'2016)            & 2.59 \textcolor{gaingreen}{\scriptsize +15.1\%} & 4.27 \textcolor{gaingreen}{\scriptsize +2.2\%} & 3.36 \textcolor{gaingreen}{\scriptsize +83.6\%} & 2.63 \textcolor{gaingreen}{\scriptsize +70.8\%} & 3.14 \textcolor{gaingreen}{\scriptsize +20.3\%} & 2.86 \textcolor{gaingreen}{\scriptsize +7.1\%} & 3.15 \textcolor{gaingreen}{\scriptsize +15.8\%} \\
    HEVC-RExt-Intra (JBHI'2017)            & 2.83 \textcolor{gaingreen}{\scriptsize +25.8\%} & 4.44 \textcolor{gaingreen}{\scriptsize +6.2\%} & 4.27 \textcolor{gaingreen}{\scriptsize +133.3\%} & 4.55 \textcolor{gaingreen}{\scriptsize +195.5\%} & 3.91 \textcolor{gaingreen}{\scriptsize +48.1\%} & 3.16 \textcolor{gaingreen}{\scriptsize +18.4\%} & 3.35 \textcolor{gaingreen}{\scriptsize +23.2\%} \\
    JPEG-XL (2019)         & 2.42 \textcolor{gaingreen}{\scriptsize +7.6\%}  & 4.34 \textcolor{gaingreen}{\scriptsize +3.8\%}  & 3.43 \textcolor{gaingreen}{\scriptsize +87.4\%}  & 2.61 \textcolor{gaingreen}{\scriptsize +69.5\%}  & 3.06 \textcolor{gaingreen}{\scriptsize +17.2\%}  & 2.85 \textcolor{gaingreen}{\scriptsize +6.7\%}   & 3.13 \textcolor{gaingreen}{\scriptsize +15.1\%}  \\
    \midrule
    L3C (CVPR'2019)             & 3.05 \textcolor{gaingreen}{\scriptsize +35.6\%} & 5.11 \textcolor{gaingreen}{\scriptsize +22.3\%} & 4.21 \textcolor{gaingreen}{\scriptsize +130.1\%} & 3.63 \textcolor{gaingreen}{\scriptsize +135.7\%} & 4.98 \textcolor{gaingreen}{\scriptsize +90.8\%} & 3.21 \textcolor{gaingreen}{\scriptsize +20.2\%} & 3.38 \textcolor{gaingreen}{\scriptsize +24.3\%} \\
    L-Infinite (CVPR'2021)      & 2.73 \textcolor{gaingreen}{\scriptsize +21.3\%} & 4.80 \textcolor{gaingreen}{\scriptsize +14.8\%} & 4.02 \textcolor{gaingreen}{\scriptsize +119.7\%} & 3.52 \textcolor{gaingreen}{\scriptsize +128.6\%} & 4.71 \textcolor{gaingreen}{\scriptsize +80.5\%} & 3.09 \textcolor{gaingreen}{\scriptsize +15.7\%} & 3.20 \textcolor{gaingreen}{\scriptsize +17.6\%} \\
    VVC-Intra (TCSVT'2021)      & 2.73 \textcolor{gaingreen}{\scriptsize +21.3\%} & 4.80 \textcolor{gaingreen}{\scriptsize +14.8\%} & 4.02 \textcolor{gaingreen}{\scriptsize +119.7\%} & 3.52 \textcolor{gaingreen}{\scriptsize +128.6\%} & 4.71 \textcolor{gaingreen}{\scriptsize +80.5\%} & 3.09 \textcolor{gaingreen}{\scriptsize +15.7\%} & 3.20 \textcolor{gaingreen}{\scriptsize +17.6\%} \\
    LC-FDNet (CVPR'2022)        & 2.40 \textcolor{gaingreen}{\scriptsize +6.7\%}  & 4.27 \textcolor{gaingreen}{\scriptsize +2.2\%}  & 3.21 \textcolor{gaingreen}{\scriptsize +75.4\%}  & 2.39 \textcolor{gaingreen}{\scriptsize +55.2\%}  & 2.84 \textcolor{gaingreen}{\scriptsize +8.8\%}   & 2.78 \textcolor{gaingreen}{\scriptsize +4.1\%}   & 2.91 \textcolor{gaingreen}{\scriptsize +7.0\%}   \\
    LC-FDNet++ (CVPR'2023)      & 2.44 \textcolor{gaingreen}{\scriptsize +8.4\%}  & 4.24 \textcolor{gaingreen}{\scriptsize +1.4\%}  & 1.96 \textcolor{gaingreen}{\scriptsize +7.1\%}   & 1.63 \textcolor{gaingreen}{\scriptsize +5.8\%}   & 2.75 \textcolor{gaingreen}{\scriptsize +5.4\%}   & 2.83 \textcolor{gaingreen}{\scriptsize +6.0\%}   & 2.87 \textcolor{gaingreen}{\scriptsize +5.5\%}   \\
    BD-VILC-Intra (TIP'2024)      & 2.42 \textcolor{gaingreen}{\scriptsize +7.5\%}  & \underline{4.20} \textcolor{gaingreen}{\scriptsize +0.5\%}  & \underline{1.93} \textcolor{gaingreen}{\scriptsize +5.5\%}   & \underline{1.61} \textcolor{gaingreen}{\scriptsize +4.5\%}   & \underline{2.71} \textcolor{gaingreen}{\scriptsize +2.7\%}   & 2.81 \textcolor{gaingreen}{\scriptsize +5.2\%}   & \underline{2.84} \textcolor{gaingreen}{\scriptsize +4.4\%}   \\
    DLPR (TPAMI'2024)            & \underline{2.34} \textcolor{gaingreen}{\scriptsize +4.0\%}  & 4.21 \textcolor{gaingreen}{\scriptsize +0.7\%}  & 3.19 \textcolor{gaingreen}{\scriptsize +74.3\%}  & 2.27 \textcolor{gaingreen}{\scriptsize +47.4\%}  & 2.77 \textcolor{gaingreen}{\scriptsize +6.1\%}   & \underline{2.76} \textcolor{gaingreen}{\scriptsize +3.4\%}   & 2.88 \textcolor{gaingreen}{\scriptsize +5.9\%}   \\
    \midrule
    Delétang et al. (ICLR'2024)         & 3.02 \textcolor{gaingreen}{\scriptsize +34.2\%} & 4.48 \textcolor{gaingreen}{\scriptsize +7.2\%}  & 4.23 \textcolor{gaingreen}{\scriptsize +131.2\%} & 3.31 \textcolor{gaingreen}{\scriptsize +115.0\%} & 3.17 \textcolor{gaingreen}{\scriptsize +21.5\%} & 2.93 \textcolor{gaingreen}{\scriptsize +9.7\%}  & 3.24 \textcolor{gaingreen}{\scriptsize +19.1\%} \\
    
    \rowcolor{lightgray}\textbf{Ours}    & \textbf{2.25} & \textbf{4.18} & \textbf{1.83} & \textbf{1.54} & \textbf{2.64} & \textbf{2.67} & \textbf{2.72} \\
    \bottomrule
    \end{tabular}
    \label{SOTACompare}
\end{table*}

\begin{table}[!t]
\renewcommand\arraystretch{1.1}
\begin{center}
\caption{Compression ratio comparison on representative datasets. The compression ratio is computed as the ratio of the original image size to the compressed image size.}
\label{tab:cr_results}
\begin{tabular}{cccc}
\toprule
Codec & C-19-R $\uparrow$ & C-19-C $\uparrow$ & L-Spine $\uparrow$ \\
\midrule
PNG (1997) & 2.58 & 1.75 & 2.27 \\
JPEG-LS (TIP'2000) & 2.91 & 1.81 & 2.06 \\
JPEG2000 (2000) & 2.89 & 1.79 & 1.92 \\
WebP (2015) & 2.87 & 1.84 & 2.42 \\
FLIF (ICIP'2016) & 3.09 & 1.87 & 2.55 \\
HEVC-RExt-Intra (JBHI'2017) & 2.83 & 1.80 & 2.05 \\
JPEG-XL (2019)  & 3.31 & 1.84 & 2.61 \\
L3C (CVPR'2019) & 2.62 & 1.57 & 1.61 \\
L-Infinite (CVPR'2021)  & 2.93 & 1.67 & 1.70 \\
VVC-Intra (TCSVT'2021) & 2.93 & 1.67 & 1.70 \\
LC-FDNet (CVPR'2022) & 3.33 & 1.87 & 2.82 \\
LC-FDNet++ (CVPR'2023) & 3.28 & 1.89 & 2.91 \\
BD-VILC-Intra (TIP'2024) & 3.31 & 1.90 & 2.95 \\
DLPR (TPAMI'2024) & 3.42 & 1.90 & 2.89 \\
Delétang et al. (ICLR'2024) & 2.65 & 1.79 & 2.52 \\
\textbf{Ours} & \textbf{3.56} & \textbf{1.91} & \textbf{3.03} \\
\bottomrule
\end{tabular}
\end{center}
\end{table}

\section{Experimental Results}
\subsection{Experimental Settings}
\subsubsection{Datasets} Following ~\cite{10402695}, we use the Flickr2k\footnote{https://github.com/limbee/NTIRE2017.} as the training dataset to fine-tune the LLM, where all images are uniformly center-cropped into 128$\times$128 patches. For comprehensive evaluation of lossless compression performance, we adopt diverse medical image datasets spanning multiple modalities and anatomical regions. Detailed statistics of these datasets are provided in Table \ref{tb:dataset_details}.

\subsubsection{Implementation Details} 
In this paper, we utilize the pre-trained Qwen2.5-7B\footnote{https://huggingface.co/Qwen/Qwen2.5-7B.} base model as the default LLM. A-LoRA is applied to fine-tune the LLM. Similar to vanilla LoRA, the A-LoRA also approximates a large matrix by two low-rank decomposed matrices, with performance governed by rank $r$ and the scaling factor $\alpha$. We ablate the rank in some predefined values, and the alpha coefficient is twice as much as the rank for a defaulted setting. After ablation studies, we set $r$ and $\alpha$ as 64, and 128, respectively (see ablation studies in Section \ref{A-LoRA}). The local modality is then cropped into patches of size 16$\times$16 (see ablation studies in Section \ref{patchsize}), which serves as inputs to the LLM. We use the randomly but differently generated binary sequences to represent different patient identity messages $\mathbf{M}$. We train the entire framework using the AdamW optimizer \cite{loshchilov2017decoupled} with an initial learning rate of $1\times10^{-4}$ and decreases by cosine decay learning scheduler \cite{loshchilov2016sgdr}. Our method is implemented using the PyTorch framework and requires 3 days to train the entire model on 4 NVIDIA A100 GPUs.

\subsubsection{Baseline Codecs} To validate the effectiveness of our method, we compare it against comparative codecs, categorized into three types: \textbf{1) Classical Codecs.} These methods rely on manually designed priors and carefully crafted frameworks. We adopt some widely used classical codecs, including PNG~\cite{boutell1997png}, JPEG-LS~\cite{weinberger2000loco}, JPEG2000~\cite{marcellin2000overview}, WebP~\cite{si2015research}, FLIF~\cite{sneyers2016flif}, HEVC-RExt-Intra~\cite{parikh2017high}, and JPEG-XL~\cite{alakuijala2019jpeg}. \textbf{2) Learned Image Compression (LIC).} LIC models typically optimize the rate-distortion trade-off directly through DNNs architectures. 
Representative LIC models include: L3C~\cite{mentzer2019practical}, L-Infinite~\cite{Bai_2021_SNIC}, VVC-Intra~\cite{bross2021overview}, LC-FDNet~\cite{9878985}, LC-FDNet++~\cite{10402695}, BD-VILC-Intra~\cite{wang2024learning} and DLPR~\cite{10378746}. \textbf{3) LLM-based Compressor.} We also reproduce the LLM-based lossless image codec proposed by Delétang et al.~\cite{deletang2024language} in our experiments. Since the LLM used in their approach is not open-source, we substitute it with Qwen2.5-7B as the default model while following their other settings.

\subsubsection{Metrics}
In this study, we utilize bits per pixel (bpp) as the metric for evaluating compression ratios. Specifically, for our method, we define the complete transmitted bitstream as
\begin{equation}
\mathcal{S}_{\mathrm{total}}
=
\mathcal{S}_{\mathrm{global}}
\,\Vert\,
\mathcal{S}_{\mathrm{local}}
\,\Vert\,
\mathcal{S}_{L}
\,\Vert\,
\mathcal{S}_{B}
\,\Vert\,
\mathcal{S}_{i},
\end{equation}
where \(\mathcal{S}_{\mathrm{global}}\) and \(\mathcal{S}_{\mathrm{local}}\) denote the compressed bitstreams generated by the global-modality path and local-modality path, respectively, while \(\mathcal{S}_{L}\), \(\mathcal{S}_{B}\), and \(\mathcal{S}_{i}\) denote the encoded side information corresponding to the segment lengths \(\{L_i\}\), the bitmap \(\mathbf{B}\), and the location index \(i\), respectively. Here, \(\Vert\) denotes bitstream concatenation. Based on this complete bitstream composition, the bitrate is computed as
\begin{equation}
\mathrm{bpp}
=
\frac{\left|\mathcal{S}_{\mathrm{total}}\right|}{H\times W},
\end{equation}
where \(\left|\mathcal{S}_{\mathrm{total}}\right|\) denotes the total number of transmitted bits required for exact decoding. Therefore, all bits required for exact decoding are accounted for in a unified manner. For the baseline codecs, bpp is computed from their original compressed file sizes following their standard settings. For bpp, the lower, the better. In addition to bpp, we report the compression ratio for a more intuitive assessment. The compression ratio is defined as the ratio of the original image size to the compressed image size:
\begin{equation}
\mathrm{CR} = \frac{\mathrm{Original\ image\ size}}{\mathrm{Compressed\ image\ size}}.
\end{equation}
For an uncompressed grayscale image with bit depth $m$, the original image size is $m \times H \times W$ bits. Since the compressed image size is $|S_{total}|$, the compression ratio of our method can be equivalently computed as
\begin{equation}
\mathrm{CR} = \frac{m \times H \times W}{|S_{total}|} = \frac{m}{\mathrm{bpp}}.
\end{equation}
Therefore, a lower bpp corresponds to a higher compression ratio. For baseline codecs, the compression ratio is computed from their original compressed file sizes in the same manner.

\subsection{Experimental Results}
\subsubsection{Model Performance} As shown in Table \ref{SOTACompare}, our proposed method achieves the new state-of-the-art (SOTA) compression performance compared to previous methods. Specifically, it outperforms all other methods in seven medical image datasets across multiple imaging modalities, including CT, MRI and ultrasonoscopy. On the C-19-R and L-Spine datasets, our approach further reduces file size by 7.6\%-17.2\% compared to the best classical codec, JPEG-XL. When compared to SOTA LIC codecs such as L3C and DLPR, our approach also demonstrates superior results. For example, the bpp of DLPR is 2.34, while our method achieves 2.25, reflecting 4.0\% improvement. Additionally, in comparison with a LLM-based codec, our method  reduces the bpp from 3.02 to 2.25 on the C-19-R dataset. To provide a more intuitive interpretation of compression performance, we further report the compression ratio, defined as the ratio of the original image size to the compressed image size, in Table~\ref{tab:cr_results}. The ranking of different methods remains consistent with the bpp results, while the compression ratio presentation makes the practical compression gain easier to interpret. For example, on the C-19-R dataset, our method achieves a compression ratio of 3.56, compared with 3.42 for DLPR and 2.65 for the LLM-based codec of Delétang et al., indicating that our framework preserves more original image content per unit compressed storage. 
These results underscore that our architecture, enhanced by dual-path compression and A-LoRA, significantly improves performance in lossless image compression.

\subsubsection{Runtime and LLM Size} We present the encoding and decoding speed for some competitive methods on the C-19-R dataset in Table \ref{tb:results_time}. The encoding and decoding time are dramatically decreasing compared to the LLM-based codec \cite{deletang2024language}. This phenomenon is attributed to the dual-path compression scheme, where LLMs only need to compress the local modality. Compared to other codecs, our method (1.5B) also renders a competitive inference speed while maintaining the SOTA compression performance. Furthermore, we conduct experiments utilizing three Qwen models with varying parameters to evaluate the impact of LLM size on compression performance. The results indicate that the degradation of compression performance is not prominent  as the model size decreases, \textit{i.e.,} smaller models can still achieve acceptable performance. Based on the current runtime results, our method does not outperform highly optimized traditional codecs such as JPEG-XL or DLPR in absolute decoding latency. In particular, Table~\ref{tb:results_time} shows that our 1.5B variant requires 39.91s for decoding, which is slower than classical or compact learned codecs. However, our intended efficiency claim is relative to fully autoregressive LLM-based codecs, rather than to all existing codecs. Compared with Delétang et al., our method reduces decoding time from 287.30s to 39.91s and encoding time from 10.47s to 1.55s, while also improving compression performance from 3.02 bpp to 2.25 bpp on C-19-R. This supports the main design goal of our dual-path framework: to avoid applying the LLM to the entire image and instead restrict the autoregressive modeling to the local modality only, thereby achieving a more favorable compression–runtime trade-off.

To better assess practical deployability, we further provide the original server-side runtime analysis with deployment-oriented single-sample latency evaluation under batch\_size = 1, as reported in Table~\ref{tab:deploy_latency}. The results show that traditional codecs such as JPEG-XL remain substantially faster in lightweight decoding scenarios, with a decoding latency of only 0.13~s compared with 24.68~s for our method. Compact neural codecs such as LC-FDNet and DLPR also achieve lower absolute latency than our framework. Therefore, our method is not intended for highly resource-constrained or real-time endpoint deployment. Nevertheless, our method achieves the best compression ratio of 2.25~bpp and provides a much better compression--efficiency trade-off than the pure LLM-based compressor, reducing decoding time from 161.27~s to 24.68~s while improving bpp from 3.02 to 2.25.
We further report throughput under parallel server-side processing in Table~\ref{tab:throughput}. Our method achieves an encoding throughput of 103.2~img/s, which is comparable to DLPR (104.6~img/s) and LC-FDNet (100.0~img/s), while maintaining the best compression performance. Although its decoding throughput of 4.0~img/s remains lower than compact neural codecs, it is still substantially higher than that of the pure LLM-based compressor (0.56~img/s). These results indicate that the proposed framework is practically promising for centralized medical image archiving and secure transmission with dedicated GPU resources, rather than CPU-only or bedside real-time deployment.

\begin{table}[!t]
    \small 
    \caption{Runtime analysis and LLM size analysis on the C-19-R dataset. 8 NVIDIA A100 GPUs (\#batch\_size:16, \#subprogress: 2 per GPU) for parallel processing.}
    \begin{tabular}{cccccc}  
    \toprule
     Codec & Bpp ↓ & Params.&  Enc. Time & Dec. Time \\
         \midrule
          JPEG-XL & 2.42  &$-$  & 0.79s & 0.12s \\
          L3C & 3.05 & 5M  & 8.21s  & 7.91s \\
          LC-FDNet & 2.40 & 23.70M &1.60s  &1.60s  \\
          DLPR & 2.34 & 22.2M & 1.53s & 2.07s   \\
          \midrule
          Delétang et al. & 3.02 & 8B  & 10.47s & 287.30s  \\
          Ours (1.5B) & 2.25 & 1.5B+2M  & 1.55s  & 39.91s  \\
          Ours (3B) & 2.27 & 3B+3M & 3.21s & 64.30s \\
          Ours (7B) & 2.25 & 7B+4M & 6.37s & 141.60s  \\
         \bottomrule
    \end{tabular}
    \label{tb:results_time}
\end{table}


\begin{table}[!t]
\centering
\caption{Single-sample encoding/decoding latency comparison under deployment-oriented settings (\#batch\_size=1). Classical codecs are evaluated on a single CPU core, while neural and LLM-based methods are evaluated on a single GPU.}
\label{tab:deploy_latency}
\resizebox{\linewidth}{!}{%
\begin{tabular}{lcccc}
\toprule
Codec & Hardware & Bpp $\downarrow$ & Enc. Time $\downarrow$ & Dec. Time $\downarrow$ \\
\midrule
JPEG-XL & 1 CPU core & 2.42 & 0.82 s & 0.13 s \\
L3C & 1 GPU & 3.05 & 0.74 s & 0.69 s \\
LC-FDNet & 1 GPU & 2.40 & 0.28 s & 0.26 s \\
DLPR & 1 GPU & 2.34 & 0.33 s & 0.41 s \\
Delétang et al. & 1 GPU & 3.02 & 5.84 s & 161.27 s \\
Ours (1.5B) & 1 GPU & 2.25 & 0.91 s & 24.68 s \\
\bottomrule
\end{tabular}%
}
\end{table}

\begin{table}[t]
\centering
\caption{Throughput comparison under parallel server-side processing. Results are reported on 8 NVIDIA A100 GPUs with batch processing.}
\label{tab:throughput}
\resizebox{\linewidth}{!}{%
\begin{tabular}{lccc}
\toprule
Codec & Bpp $\downarrow$ & Encoding Throughput $\uparrow$ & Decoding Throughput $\uparrow$ \\
\midrule
L3C & 3.05 & 19.5 img/s & 20.2 img/s \\
LC-FDNet & 2.40 & 100.0 img/s & 100.0 img/s \\
DLPR & 2.34 & 104.6 img/s & 77.3 img/s \\
Delétang et al. & 3.02 & 15.3 img/s & 0.56 img/s \\
Ours (1.5B) & 2.25 & 103.2 img/s & 4.0 img/s \\
\bottomrule
\end{tabular}%
}
\end{table}
\subsubsection{Security Validation and Visualization of Stego Images} 
\begin{figure}[!t]
 \centering
 \includegraphics[width=\linewidth]{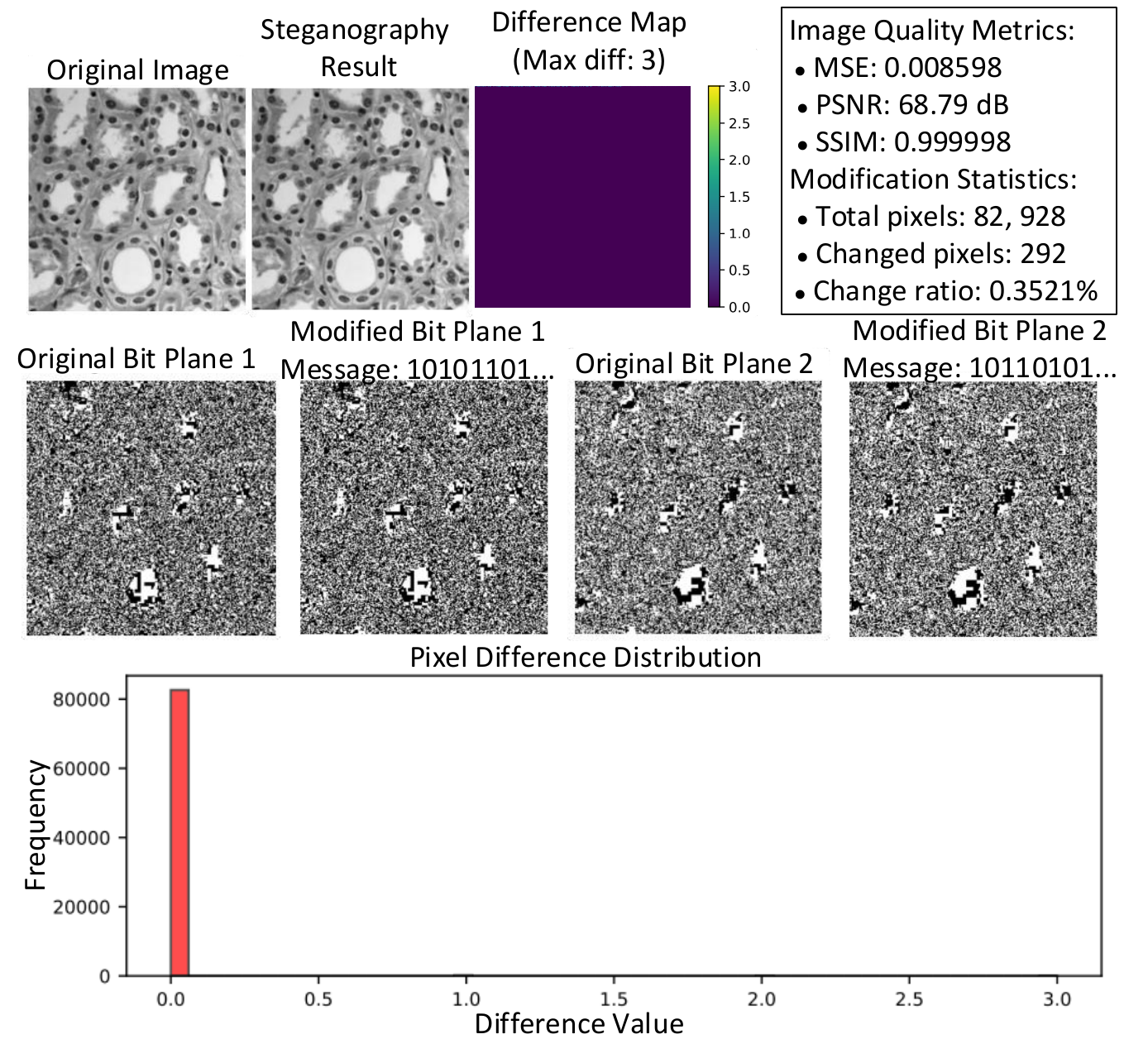}
 \caption{Visualization of stego images. The proposed segmented message steganography algorithm achieves an invisible message encryption.}
\label{fig:stego}
\end{figure}

As shown in Figure \ref{fig:stego}, our proposed segmented message
steganography algorithm achieves a remarkable invisibility to human observers, as the visual fidelity metrics are excellent, \textit{e.g.,} PSNR of 68.79 dB (very high - above 40 dB is considered excellent), and SSIM of 0.999998 (nearly perfect structural similarity)~\cite{kaissis2020secure}. 
Meanwhile, only 292 pixels out of 82,928 total pixels are changed (0.3521\% change ratio), indicating efficient embedding with minimal alterations. The histogram of pixel difference distribution has a very high peak at 0, indicating that most of the pixels have not been modified.
\begin{figure*}[t]
 \centering
 \includegraphics[scale=0.52]{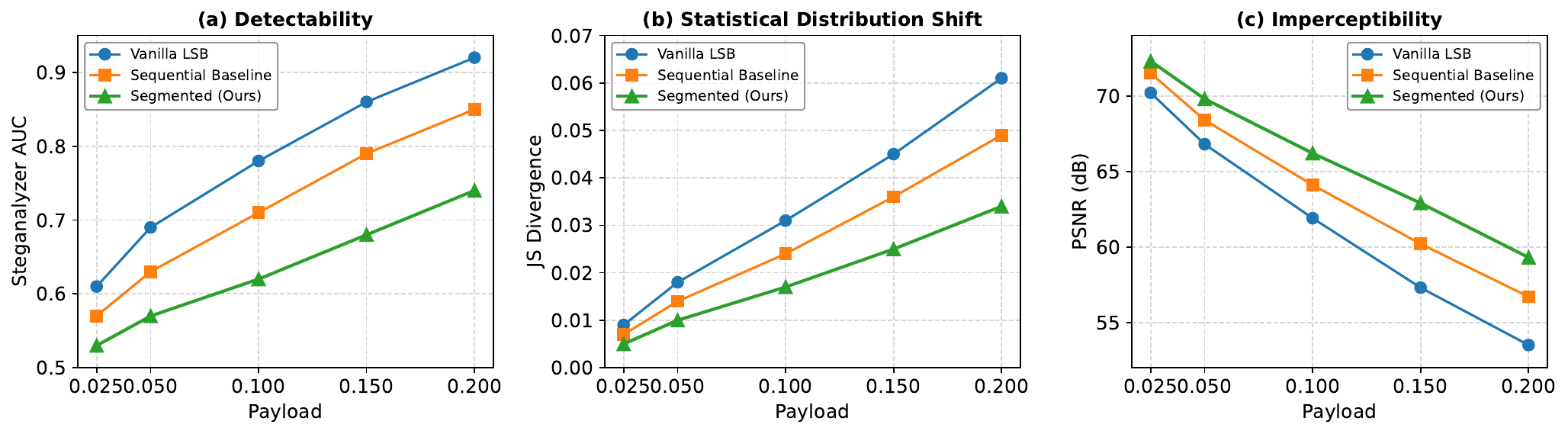}
\caption{Security analysis of different steganographic strategies under matched payload settings: vanilla LSB replacement, sequential local-plane embedding, and our segmented embedding strategy. The sequential local-plane embedding baseline uses the same adaptive modalities decomposition as our method, but embeds the secret message as a continuous bitstream into the low-significance local modality in a lower-to-higher plane order, without explicit plane-wise message segmentation. (a) Detectability measured by the AUC of SR-Net. (b) Statistical distribution shift measured by JS divergence between cover and stego images. (c) Imperceptibility measured by PSNR. Lower AUC and JS divergence, and higher PSNR, indicate better concealment performance. Note that the payload bpp used here is different from the compression bpp reported for coding efficiency: the former measures embedding capacity, while the latter measures the number of bits required to represent the compressed image.}
\label{fig:secu}
\end{figure*}
\begin{table*}[t]
\centering
\caption{Quantitative security analysis of different steganographic strategies on the C-19-R dataset under matched payload settings. Detectability is measured by the AUC of SR-Net. Lower AUC and JS divergence, and higher PSNR, indicate better concealment performance.}
\label{tab:stego_security}
\setlength{\tabcolsep}{5pt}
\renewcommand{\arraystretch}{1.1}
\begin{tabular}{c|ccc|ccc|ccc}
\hline
\multirow{2}{*}{Payload} 
& \multicolumn{3}{c|}{SR-Net AUC $\downarrow$}
& \multicolumn{3}{c|}{JS Divergence $\downarrow$}
& \multicolumn{3}{c}{PSNR (dB) $\uparrow$} \\
\cline{2-10}
& Vanilla LSB & Sequential Baseline & Ours
& Vanilla LSB & Sequential Baseline & Ours
& Vanilla LSB & Sequential Baseline & Ours \\
\hline
0.025 & 0.63 & 0.56 & \textbf{0.52}
      & 0.008 & 0.006 & \textbf{0.005}
      & 70.2 & 71.5 & \textbf{72.2} \\
0.050 & 0.69 & 0.60 & \textbf{0.56}
      & 0.018 & 0.014 & \textbf{0.010}
      & 66.8 & 68.1 & \textbf{69.3} \\
0.100 & 0.78 & 0.70 & \textbf{0.61}
      & 0.031 & 0.026 & \textbf{0.017}
      & 61.9 & 64.2 & \textbf{66.3} \\
0.150 & 0.86 & 0.77 & \textbf{0.67}
      & 0.045 & 0.036 & \textbf{0.025}
      & 57.3 & 60.2 & \textbf{62.9} \\
0.200 & 0.92 & 0.83 & \textbf{0.73}
      & 0.061 & 0.049 & \textbf{0.034}
      & 53.5 & 56.8 & \textbf{59.3} \\
\hline
\end{tabular}
\end{table*}
In the steganographic analysis, the payload measured in bpp refers to the average number of secret bits embedded into each image pixel, rather than the compression bitrate used to evaluate coding efficiency. Specifically, given a secret message of length $L$ bits and a cover image of spatial size $H\times W$, the payload is defined as $\mathrm{Payload}~(\mathrm{bpp}) = L / (H \times W)$. Therefore, different payload settings correspond to different embedding capacities. In this work, we vary the payload from 0.025 to 0.2 bpp, and conduct the security evaluation on the C-19-R dataset under matched embedding-capacity settings. As shown in Figure~\ref{fig:secu} and Table~\ref{tab:stego_security}, we evaluate the proposed segmented message steganography from three complementary perspectives, including detectability, statistical distribution shift, and visual imperceptibility. In Figure~\ref{fig:secu}(a), detectability is assessed using the AUC of SR-Net~\cite{boroumand2018deep}, a representative deep steganalyzer for image steganalysis, where a lower AUC indicates that the stego image is harder to distinguish from the cover image. Quantitatively, our segmented strategy consistently achieves the lowest AUC across all payloads. For example, at 0.2 bpp, the AUC is reduced to about 0.73, compared with about 0.83 for the sequential baseline and about 0.92 for vanilla LSB. Here, sequential baseline denotes a baseline that uses the same adaptive modalities decomposition as our framework and restricts message embedding to the low-significance local modality $\mathbf{x}^{1:s}$. Different from our segmented strategy, it does not explicitly split the secret message into plane-wise sub-messages. Instead, the message is treated as a single continuous bitstream and sequentially embedded bit-by-bit into the local bit planes following a fixed lower-to-higher plane order. Even at a moderate payload of 0.1 bpp, our method achieves an AUC of about 0.61, lower than that of the sequential baseline (about 0.70) and vanilla LSB (about 0.78). This indicates that the proposed method is more resistant to steganalytic detection, especially in the high-payload regime. In Figure~\ref{fig:secu}(b), the JS divergence between the cover-image distribution and the stego-image distribution is reported. A lower JS divergence means that the embedding process introduces less statistical deviation from the original image distribution. Our method again yields the smallest JS divergence across all payloads. For instance, at 0.2 bpp, the JS divergence is about 0.034, compared with about 0.049 for the sequential baseline and about 0.061 for vanilla LSB. At 0.1 bpp, the JS divergence is about 0.017, compared with about 0.026 and 0.031 for the sequential baseline and vanilla LSB, respectively. These results indicate that our segmented embedding better preserves the natural statistical characteristics of medical images. In Figure~\ref{fig:secu}(c), visual fidelity is measured by PSNR, where a higher value corresponds to lower distortion. The proposed method maintains the highest PSNR over the full payload range. Specifically, at 0.2 bpp, our method still preserves about 59.3 dB, which is clearly higher than about 56.8 dB for the sequential baseline and about 53.5 dB for vanilla LSB. At 0.1 bpp, the corresponding PSNR values are about 66.3 dB, 64.2 dB, and 61.9 dB, respectively. These results jointly show that, by embedding the secret message only into the local modality and distributing it across multiple bit planes, the proposed segmented strategy achieves a more favorable trade-off among SR-Net detectability, statistical consistency, and image fidelity than direct bit-replacement schemes.

\subsection{Ablation Studies \label{subsec:ablation_studies}}
To further analyze our architecture, we conduct ablation studies on the C-19-R dataset.
\subsubsection{Component Ablation Studies}  
\begin{figure}[t]
 \centering
 \includegraphics[width=\linewidth]{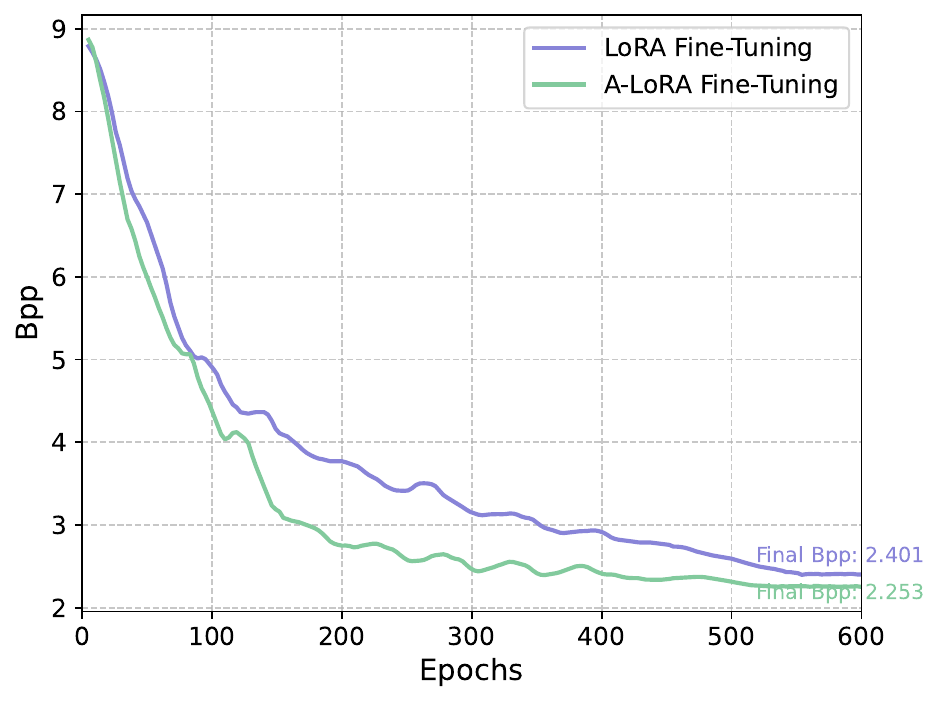}
 \caption{The bpp score tendency of our method with different fine-tuning strategy during the training period.}
\label{fig:inc_curves}
\end{figure}

To further validate the contribution of each key component in our framework, we conduct component-wise ablation studies on the C-19-R dataset. Table~\ref{component} presents a progressive component-wise ablation to explicitly evaluate the contributions of the Dual-Path design, steganography strategy, visual prompt, and fine-tuning scheme. We first establish a VAE-Only baseline, where the Dual-Path compression is disabled and the entire image is compressed using only the VAE branch. In this setting, the image is not decomposed into global and local modalities, and neither the LLM-based local compression path nor the steganography module is involved. This baseline produces 4.45 bpp. Enabling the proposed Dual-Path design reduces the bitrate from 4.45 to 4.09 bpp, demonstrating that decomposing the image into global and local modalities is beneficial for lossless medical image compression. This validates the core motivation of our framework: the VAE branch efficiently handles the global modality, whereas the LLM branch is better suited to model local fine-grained details.
Next, we examine the contribution of the steganography strategy within the Dual-Path framework. Adding vanilla LSB steganography reduces the bpp to 3.77, while adopting the proposed segmented steganography further reduces it to 3.34. The clear improvement over vanilla LSB shows that the proposed segmented strategy is not only more suitable for secure embedding, but also more compatible with the compression pipeline.
We then introduce the visual prompt, \textit{i.e.,} the embeddings of the global modality used to guide the LLM-based local compression path. This further reduces the bpp from 3.34 to 2.77, indicating that visual prompts effectively inject image-aware priors into the pretrained LLM and substantially improve local-modality compression.
Finally, we compare two fine-tuning strategies on top of visual prompting. Vanilla LoRA decreases the bpp from 2.77 to 2.40, while the proposed A-LoRA achieves the best result of 2.25 bpp. As shown in Figure \ref{fig:inc_curves}, it is evident that utilizing A-LoRA fine-tuning strategy rather than vanilla LoRA further improves the compression performance and accelerates the training speed. This validates that anatomical priors provide a more suitable adaptation initialization for medical image compression than standard LoRA. These results demonstrate that all components are indispensable for the proposed joint lossless compression and steganography framework.

\begin{table*}[!t]
\small
\centering
\caption{Component-wise ablation results on the C-19-R dataset. We explicitly analyze the contributions of the dual-path design, steganography strategy, visual prompt, and A-LoRA. Lower bpp indicates better compression performance.}
\label{tab:component}
\begin{tabular*}{\textwidth}{@{\extracolsep{\fill}}cccccccc@{}}
\toprule
\multirow{2}{*}{Baseline} & \multirow{2}{*}{Dual-Path} & \multicolumn{2}{c}{Steganography} & \multirow{2}{*}{Visual Prompt} & \multirow{2}{*}{LoRA} & \multirow{2}{*}{A-LoRA} & \multirow{2}{*}{Bpp $\downarrow$} \\
\cmidrule(lr){3-4}
& & Vanilla LSB & Segmented Steg. & & & & \\
\midrule
\ding{51} & \ding{55} & \ding{55} & \ding{55} & \ding{55} & \ding{55} & \ding{55} & 4.45 \\
\ding{51} & \ding{51} & \ding{55} & \ding{55} & \ding{55} & \ding{55} & \ding{55} & 4.09 \\
\ding{51} & \ding{51} & \ding{51} & \ding{55} & \ding{55} & \ding{55} & \ding{55} & 3.77 \\
\ding{51} & \ding{51} & \ding{55} & \ding{51} & \ding{55} & \ding{55} & \ding{55} & 3.34 \\
\ding{51} & \ding{51} & \ding{55} & \ding{51} & \ding{51} & \ding{55} & \ding{55} & 2.77 \\
\ding{51} & \ding{51} & \ding{55} & \ding{51} & \ding{51} & \ding{51} & \ding{55} & 2.40 \\
\ding{51} & \ding{51} & \ding{55} & \ding{51} & \ding{51} & \ding{55} & \ding{51} & 2.25 \\
\bottomrule
\end{tabular*}
\label{component}
\end{table*}

\subsubsection{Choice of Retention hyperparameter $\beta$} 
\label{retention}
The retention hyperparameter $\beta$ controls the proportion of information allocated to global modalities during the adaptive modalities decomposition process. To determine its optimal value, we conduct experiments on the C-19-R dataset, evaluating the compression performance for $\beta \in \{0.6, 0.7, 0.8, 0.9\}$. The results are presented in Table \ref{parameter}. The results demonstrate that $\beta = 0.8$ provides the optimal trade-off between global and local information retention, achieving the best compression performance. Smaller values of $\beta$ (\textit{e.g.,} 0.6, 0.7) underutilize global information, while larger value (\textit{e.g.,} 0.9) compromises the effectiveness of local modality compression. Therefore, we set $\beta = 0.8$ as the default value in our method.
\begin{table}[!t]
\small
\caption{Impact of $\beta$ on Compression Performance.}
\begin{center}
\begin{tabular*}{0.42\textwidth}{@{\extracolsep{\fill}}cccccl@{}}
\toprule
$\beta$  & 0.6 & 0.7 & 0.8 & 0.9 \\ \hline
Bpp ↓ & 2.52 & 2.41 & \textbf{2.25} & 2.38\\
\bottomrule
\end{tabular*}
\label{parameter}
\end{center}
\end{table}

\subsubsection{Different Rank $r$ for A-LoRA} 
\label{A-LoRA}
The rank and corresponding alpha coefficient in A-LoRA would significantly affect the compression performance. As shown in Table \ref{LoRAsize}, we ablate the rank in some predefined values, and alpha is twice as much as the rank for a defaulted setting. Theoretically, larger rank leads to more powerful representation ability, resulting in better compression ratios. However, as shown in Table \ref{LoRAsize}, the compression performance remains essentially unchanged as the rank increases beyond a certain point. To strike a great balance between performance and efficiency, we set the rank and alpha to 64 and 128, respectively.
\begin{table}[!t]
\small
\begin{center}
\caption{Ablation experiments for A-LoRA, test results on the C-19-R dataset, using bpp as metric.}
\begin{tabular*}{0.47\textwidth}{@{\extracolsep{\fill}}cccc@{}}
\toprule
Rank $r$ & Alpha & Bpp ↓ &Gain \\  \hline
8 &16  & 2.30  &+2.2\%\\
16 &32  & 2.29 &+1.78\%\\ 
32 &64  & 2.27 &+0.89\%\\ 
64 &128  & \textbf{2.25} & $-$\\ 
128 &256  & 2.26 &+0.44\%\\
\bottomrule
\end{tabular*}
\label{LoRAsize}
\end{center}
\end{table}

\subsubsection{Patch Size} 
\label{patchsize}
The patch size determines the context length of LLM, which can affect the compression performance. Table \ref{patchszize} reveals that moderate patch size, \textit{i.e.,} 16 $\times$ 16 achieves the optimal performance, as short-context length provides limited contextual information for LLM while long-context impairs the performance due to the degradation of position embeddings.
\begin{table}[!t]
\small
\caption{Comparison of different patch sizes for LLM. Moderate patch size is the best.}
\begin{center}
\begin{tabular*}{0.47\textwidth}{@{\extracolsep{\fill}}cccccc@{}}
\toprule
Patch Size  & $8 \times 8$ & $12 \times 12$ & $16 \times 16$ & $24 \times 24$ & $32 \times 32$\textbf{}\\  \hline
Bpp ↓ & 2.74 & 2.47 & \textbf{2.25} & 2.32 & 2.91\\
\bottomrule
\end{tabular*}
\label{patchszize}
\end{center}
\end{table}

\section{Conclusion and Future Work}
In this paper, we propose the joint lossless compression and steganography framework, which ensures the security for compressing medical images. Under this framework, the adaptive modalities decomposition strategy is first designed to explicitly decompose images into global and local modalities. Then, we present a dual-path compression scheme, which takes the first attempt to deploy LLMs’ unprecedented intelligence for arithmetic coding in the combined paradigm. Combined with the carefully designed anatomical priors-based low-rank adaptation (A-LoRA) fine-tuning strategy, our method achieves superior compression performance against other SOTA codecs with comparable inference time. As this framework is still in its early stages, future works could focus on developing more lightweight LLMs for compression, which makes it feasible to deploy the method in resource-constrained medical scenarios. 

\printbibliography
\end{document}